\documentclass[preprint,10pt]{aastex}
\usepackage{amssymb,latexsym,amscd,amsmath,graphicx,ifpdf}
\usepackage{longtable}
\pdfoutput=1

\ifpdf

\DeclareGraphicsExtensions{.pdf}
\def\be{\begin{equation}}
\def\ee{\end{equation}}
\def\ba{\begin{array}}
\def\ea{\end{array}}
\def\bea{\begin{eqnarray}}
\def\eea{\end{eqnarray}}
\def\bi{\begin{itemize}}
\def\ei{\end{itemize}}

\def\half{{\textstyle{1\over2}}}

\def\ve{\varepsilon}

\shorttitle{Neutron star inner crust survey with CLDM}
\shortauthors{Newton et al.}
\begin{document}
\title{A survey of the parameter space of the compressible liquid drop model as applied to the neutron star inner crust}

\author{W.~G.~Newton}
\affil{Department of Physics and Astronomy, Texas A\&M University-Commerce, Commerce, Texas 75429-3011, USA}
\email{william.newton@tamuc.edu}
\author{M.~Gearheart}
\affil{Department of Physics and Astronomy, Texas A\&M University-Commerce, Commerce, Texas 75429-3011, USA}
\and
\author{Bao-An~Li}
\affil{Department of Physics and Astronomy, Texas A\&M University-Commerce, Commerce, Texas 75429-3011, USA}
\date{\today}

\begin{abstract}


We present a systematic survey the range of predictions of the neutron star inner crust composition, crust-core transition densities and pressures, and density range of the nuclear `pasta' phases at the bottom of the crust provided by the compressible liquid drop model in the light of current experimental and theoretical constraints on model parameters. Using a Skyrme-like model for nuclear matter, we construct baseline sequences of crust models by consistently varying the density dependence of the bulk symmetry energy at nuclear saturation density, $L$, under two conditions: (i) that the magnitude of the symmetry energy at saturation density $J$ is held constant, and (ii) $J$ correlates with $L$ under the constraint that the pure neutron matter (PNM) EoS satisfies the results of ab-initio calculations at low densities. Such baseline crust models facilitate consistent exploration of the $L$ dependence of crustal properties. The remaining surface energy and symmetric nuclear matter parameters are systematically varied around the baseline, and different functional forms of the PNM EoS at sub-saturation densities implemented, to estimate theoretical `error bars' for the baseline predictions. Inner crust composition and transition densities are shown to be most sensitive to the surface energy \emph{at very low proton fractions} and to the behavior of the sub-saturation PNM EoS. Recent calculations of the energies of neutron drops suggest that the low-proton-fraction surface energy might be higher than predicted in Skyrme-like models, which our study suggests may result in a greatly reduced volume of pasta in the crust than conventionally predicted.

\end{abstract}


\maketitle

\section{\label{sec1}Introduction}

The increasing wealth and sensitivity of observations associated with neutron star systems is demanding ever more sophisticated theoretical models of their structure and dynamics from micro- to macro-scopic scales. Many proposed theoretical models for observed phenomena, such as pulsar glitches, or potentially observable phenomena, such as gravitational waves from stellar oscillations or deformations of the star, involve incorporating global properties of the star, properties of the elastic crust, and the coupling between the two \cite{Link1999, Gearheart2011,Sotani2011,Lindblom2000, Wu2001,Glampedakis2006, Wen2011}.

One goal of neutron star physics is to make use of the available observations to provide data complementary to those provided by terrestrial nuclear experiments in order to better understand the microphysics of dense matter \cite{Lattimer2001}. In order to extract microphysical constraints from observations, the theoretical description of the observations should ideally use consistent models of the nuclear physics throughout the star. Simple global properties of a neutron star such as its mass, radius and total moment of inertia are relatively insensitive to the properties of the crust itself, which contributes a small fraction of the total to each; in such cases the dominant theoretical ingredient is the uniform nuclear matter equation of state (EoS) for the core \cite{Urbanec2005}. However, in models where the crust-core interplay is important there will be a sensitivity to both global (and possibly microscopic) core and crust properties. The dominant uncertainties in nuclear matter EoSs stem from the uncertainties in the nuclear symmetry energy as a function of density - the function that encodes the energy cost of making neutron star matter more neutron-rich. The dependence on the symmetry energy of several global stellar properties such as the radius, and crustal properties such as crust thickness, has been well studied. However, little attention has been paid to the dependence on the symmetry energy of more complex properties that depend both on crust and core properties. This requires crust and core models derived using consistent symmetry energy behaviors and spanning the range of uncertainty in symmetry energy. For the most part, only a few crust models representing specific cases of symmetry energy behavior have been used. The most widely used crust models employ the compressible liquid drop model (CLDM), and its computational simplicity lends itself to calculation of a large number of crust models. Therefore this paper aims to survey the predictions of this class of crust models over its parameter space constrained by the most recent nuclear theoretical and experimental developments, and to provide a set of crust EoSs spanning the model's uncertainties to facilitate more consistent studies of the nuclear physics dependence of neutron star phenomena.  Our aim is to do so in a way unbiased by a particular theoretical method for extracting the CLDM parameters, but rather to attempt to take into account theoretical uncertainties from all such methods in a simple way.

The outer and inner layers of the neutron star crust are defined by the absence and presence of a neutron fluid external to the nuclear clusters respectively. Additionally, a third layer is often predicted to be present at the bottom of the inner crust, consisting of exotic nuclear geometries collectively termed nuclear `pasta'; this layer may be called the mantle \cite{Gusakov2004}. The mantle can contain two distinct sub-layers: (1) the layer in which the pasta structures contain nuclear matter of non-zero proton fraction, surrounded by a fluid of pure neutron matter (`normal pasta'), and (2) the inverse situation in which the pasta structures are bubbles containing pure neutron matter, surrounded by a higher density fluid of nuclear matter with a non-zero proton fraction (`bubble pasta'). In the latter layer, protons themselves are delocalized. Whether these layers are to be included when modeling the bottom of the inner crust and the transition to the neutron star core will have important consequences for the physics of the crust-core interface.

There are a number of important crustal quantities that are sensitive to the symmetry energy of the nuclear EoS around saturation density \cite{Oyamatsu2007,Steiner2008}. The behavior of the symmetry energy at this density is usefully characterized by its magnitude at saturation density $J$, and its slope with respect to density $L$. The crust-core transition density has been shown to be sensitive to $L$ \cite{Oyamatsu2007,Horowitz2001,Xu2009}. The crust thickness, mass and moment of inertia, important in the modeling of pulsar glitches \cite{Link1999}, is sensitive to the transition pressure, which has been shown to correlate with the slope and curvature of the symmetry energy at densities below saturation \cite{Ducoin2011}. The maximum and typical size of mountains on NS crusts \cite{Ushomirsky2000, Haskell2006} and frequencies of torsional crustal oscillations \cite{Piro2005, SteinerWatts2009,Samuelsson2007, Andersson2009} depends on the thickness of the elastically rigid part of the crust and its shear modulus, which depends on the characteristic quantities which define the lattice: the inter-ion spacing $a$ and nuclear charge $Z$ \cite{Ogata1990,Strohmayer1991,Chugunov2010,Gearheart2011}. Various transport properties within the crust will also depend somewhat on the crustal composition through the densities of the various components, the electron fraction and the lattice spacing. The region of  densities at which the neutrons are predicted to form a superfluid via the $^1$S$_0$ channel has a model dependence; the possibility of layers of the crust in which the neutron fluid is normal exists, and will depend on the crust-core transition density, and the existence of a neutron superfluid in the pasta layers will also depend on their extent. Also, the melting temperature of the crustal lattice depends on the compositional parameters, and its magnitude relative to the superfluid transition density will impact the dynamics of the crust.

The `pasta' phases \cite{Ravenhall1983, Oyamatsu1984} might have very different behaviors to the purely crystalline phase. The shear modulus of the pasta phases is still very uncertain, and will have a large effect on the crust-core boundary layer, of particular importance to the damping of r-mode oscillations \cite{Lindblom2000,Wen2011} and the spectrum of torsional oscillations \cite{Gearheart2011}. The momentum conservation requirement for the direct Urca neutrino cooling process to be allowed might be satisfied in the bubble phases due to nucleon scattering off the non-spherical nuclei \cite{Gusakov2004}. Thermodynamic and neutrino scattering properties of the pasta phases could also be rather different \cite{Watanabe2000, Watanabe2000_2, Watanabe2003}. The extent of the pasta phases in the crust will determine the importance of those phases to crustal processes and the dynamics of the crust-core coupling. The region of the inner crust occupied by the `pasta' phases is also strongly dependent on $L$, since, while the density of transition to pasta remains roughly constant in a given model \cite{Oyamatsu2007}, the crust-core transition density decreases with increasing $L$, reducing the density range of the pasta phases at higher $L$.

The compressible liquid drop model (CLDM) \cite{BBP1971} has proven to be a useful tool for the study of neutron star crust \citet{Ravenhall1972, Mackie1977, Lattimer1977, Lattimer1978, Lamb1978, Lamb1981, Lamb1983, Ravenhall1983, Pethick1983, Iida1997, Douchin2000, Watanabe2000, Watanabe2000_2, Watanabe2003, Steiner2008}. Developed from the liquid drop and droplet models of terrestrial nuclei, the CLDM focusses on the average, macroscopic properties of nuclear clusters in the crust such as mass and radius, whilst neglecting the quantum shell effects. The bulk nuclear matter interior and external neutron fluid are calculated using the same nuclear matter EoS, while the surface energy is separately parameterized. The surface energy parameters are typically determined through calculations of the interfacial energy between two regions of semi-infinite nuclear matter (SINM) or through fitting the CLDM to nuclear masses.  The results of such calculations are sensitive to the theoretical formalism of the surface energy \cite{Steiner2005} and thus introduce further model dependence on top that from the uniform nuclear matter EoS. The neutron star inner crust is assumed to have the structure of a regular lattice, and the Wigner-Seitz (WS) approximation is employed in which the unit cell, which generally will have a cubic or more complicated structure, can be replaced by a spherical cell of the same volume. The WS approximation is appropriate if the nuclear size is much smaller than the cell size, but breaks down when the two sizes become comparable. Despite its physical simplifications, the CLDM provides a good description of the average microscopic properties of the neutron star crust, including the energetically preferred nuclear mass, size, WS cell size and density of dripped neutrons. Its amenability to quick calculation and physical transparency have made it the most widely used model for the computation of the EoS and composition of the inner neutron star crust and of inhomogeneous nuclear phases in core collapse supernovae \cite{BBP1971, Douchin2001, Lattimer1991}.

Despite its wide use, the CLDM possesses several drawbacks. Nuclear shell effects, or those arising from scattering of dripped neutrons off of nuclear clusters, are not accommodated, and have the potential to significantly affect the ordering of transitions between the pasta phases \cite{Magierski2002}, as well as the equilibrium size of the nuclear clusters and transport properties such as contributions to heat transport from nuclear components \cite{Sandulescu2007} and entrainment of dripped neutrons by clusters \cite{Chamel2005b}. Secondly, the WS approximation is expected to break down when the nuclear separation becomes comparable to the cell size \cite{Chamel2007}, which occurs in the mantle. Thirdly, effects that act over ranges greater than the unit cell are not consistently accommodated in the CLDM; larger scale self-organization of pasta phases and long wavelength transport effects are all unaccounted for. Many of these problems are eliminated by more sophisticated models such as Thomas-Fermi (TF), Extended TF (ETF) and ETF + Strutinsky Integral (ETFSI) (e.g \cite{Buchler1971,Oyamatsu1993,Cheng1997,Onsi2008},  where the latter includes a self-consistent treatment of shell effects), the 1D or 3D-Hartree-Fock (HF) methods e.g. \cite{Sandulescu2007,Negele1973,Montani2004,Baldo2005,Magierski2002,Gogelein2008,Newton2009}, the latter of which relieves the spherical-WS approximation), and quantum molecular dynamics (QMD) methods \cite{Maruyama1998,Horowitz2004,Watanabe2001,Sonoda2007}. Certain of these methods are too computationally demanding for the kind of parameter survey we will present, so as a first step will shall limit ourselves to the CLDM, with a view to expanding the methodology in future.

The purpose of this paper is as follows. (1) Provide a `baseline' set of inner crust EoSs together with their associated crust compositions generated by varying the slope of the symmetry energy within limits imposed by terrestrial experiments and theoretical pure neutron matter (PNM) calculations, such that a suitable crust EoS can be found for a core EoS with the same symmetry energy properties at saturation, enabling more consistent neutron star modeling. (2) Systematically explore the remaining model dependence of the crustal composition, focussing on the experimental and theoretical uncertainties of the SNM EoS and the strength of the surface energy in the CLDM. In section~2 we outline the CLDM. In section~3 we describe our model for the bulk nuclear matter EoS, and the experimental and theoretical PNM constraints we will use. In section~4 we discuss the range of parameters used in the prescription for the surface energy. In section~5 we present the set of baseline results, before exploring the effect of varying surface energy and symmetric nuclear matter parameters as well as the functional form of the sub-saturation density symmetry energy, on the crust composition in section~6. In section~7 we discuss how the EoS and surface uncertainties can impact important crustal properties such as the mass fractions of the various layers, the shear modulus and the melting temperature, and we conclude in section~8.

\section{The Compressible Liquid Drop Model}

We use the CLDM originally formulated by BBP \cite{BBP1971} and updated to incorporate non-spherical nuclear shapes by Watanabe, Iida and Sato \cite{Watanabe2000,Watanabe2000_2,Iida1997}. For completeness, we recall the main ingredients of the CLDM. Assuming a regular crystal structure for crustal matter, a repeating unit cell can be identified in which a single nucleus (or pasta segment) resides, immersed in an external, uniform neutron fluid. A homogeneous gas of electrons is present to neutralize the positive protonic charge. We consider the three canonical geometries for nuclear clusters (the `pasta' shapes): spherical, cylindrical and planar, specified by a dimensionality parameter $d=3, 2, 1$ respectively. The particular shape can refer to the geometry of the charged nuclear phase of matter (i.e. that which contains the protons) or the pure neutron phase (the bubble phases). The Wigner-Seitz (WS) approximation is employed, in which the physical unit cell is replaced by one of equal volume with the same geometry as the nuclear cluster. The total energy density of the matter can be written (neglecting rest masses)
\be\label{eq:ecell}
	\ve_{\rm cell}(r_{\rm c},x,n,n_{\rm n}) = v \big[ n E(n,x) + \ve_{\rm exch} + \ve_{\rm thick} \big] + u(\ve_{\rm surf} + \ve_{\rm curv}) + u \ve_{\rm Coul} + (1-v) n_{\rm n} E(n_{\rm n}, 0) + \ve_{\rm e} (n_{\rm e}),
\ee

\noindent where $r_{\rm c}$ is the radius (half-width in the case of planar geometry) of the WS cell, $u = (r_{\rm N} / r_{\rm c})^d$ is the volume fraction occupied by the nuclear cluster of radius/half-width $r_{\rm N}$, $x$ and $n$ are the proton fraction and baryon density of the charged nuclear component and $n_{\rm n}$ the baryon density of the neutron fluid. $n_{\rm e}$ is the number density of electrons. Charge neutrality demands $n_{\rm e}= vnx$ where $v$ is the volume fraction of the charged nuclear component, defined as
\be\label{eq:v}
	v = \left\{
  \begin{array}{l l}
	\displaystyle u & \quad {\rm charged \; nuclear \; clusters}\\
	\displaystyle 1-u& \quad{\rm bubbles},\\
  \end{array} \right.
\ee
\noindent and the global baryon number density is related to the local baryon densities through $n_{\rm b} = vn + (1-v)n_{\rm n}$. The contributions to the energy density of the cell are as follows.

The bulk nuclear energy per particle for matter inside nuclear clusters $E(n,x)$ and the pure neutron matter outside $E(n_{\rm n},0)$ is obtained using the same model of the uniform nuclear matter EoS as outlined in the following section.

The Coulomb energy density, which includes the lattice energy (the proton-electron contribution) is given in the WS approximation by
\be\label{eq:ecl}
    \ve_{\rm (C+L)} = 2\pi (exnr_{N})^{2}f_{d}(u); \;\;\;  f_{d}(u) = \frac{1}{d+2} \left[\frac{2}{d-2}\left(1-{du^{1-2/d}\over 2} \right) + u\right].
\ee
The correction to the local energy density of the charged nuclear component from the surface thickness  and the proton Coulomb exchange energy are given by \cite{BBP1971}
\be\label{eq:ethick}
	\ve_{th}(k,x) = -\frac{4}{9}\pi e^2w^2x^2k^3n; \;\;\;\;\; \ve_{ex}(k,x) = -\frac{3}{4\pi}2^{1/3}e^2x^{4/3}kn,
\ee
\noindent where $k = (1.5\pi^2 n)^{1/3}$ and $w$ is a distance representing the surface thickness, taken to be $w \approx 0.75$ fm. The energy density of the free electron gas is taken to be simply the free Fermi gas expression
\be
    \ve_{\rm e} = \frac{3}{4}\hbar c k_{\rm e} n_{\rm e},
\ee
\noindent  where $k_{\rm e} = (3\pi^2 n_{\rm e})^{1/3}$ neglecting electron screening and the electron exchange energy, which are small corrections compared with the electron kinetic energy \cite{BBP1971}.

The surface energy density can be written in terms of the surface and curvature tensions $\sigma_{\rm s}$,$\sigma_{\rm c}$ as
\be
    \ve_{\rm surf} = \frac{d \sigma_{\rm s}}{r_{N}}; \;\;\;  \ve_{\rm curv} = \frac{d (d-1) \sigma_{\rm c}}{r_{N}^2}.
\ee

\noindent  Inspired by one- and two-fluid Thomas-Fermi theory, the first formulations of the CLDM parameterized the surface energy in terms of the proton fraction $x$, the bulk nuclear and neutron matter densities $n, n_{\rm n}$, and the difference in energy per particle between the two bulk phases \cite{BBP1971,Ravenhall1972,Mackie1977,Iida1997}. However it was noted that the surface energy should be evaluated along the curve of coexistence between the bulk nuclear and pure neutron matter \cite{Ravenhall1983.2,Lattimer1985} and therefore the quantities upon which it depends $n, x, n_{\rm n}$ are not all independent variables. At zero temperature, the surface and curvature tensions may be parameterized in terms of one variable, which is most conveniently chosen to be the proton fraction $x$. The inclusion of a neutron skin accounts for the dependence of surface properties on the bulk nuclear densities. In this work we omit the neutron skin; the effects of the skin will be examined in a subsequent study. We use the surface and curvature tension parameterization first used by Lattimer, Pethick, Ravenhall and Lamb (LPRL) \cite{Ravenhall1983.2, Lattimer1985} and subsequently generalized by Lorenz and Pethick \cite{Lorenz1991,LorenzPethick1993}:
\be \label{eqn:surf}
    \sigma_{\rm s}(x)= \sigma_0 { 2^{p+1} + b \over {1 \over x^p} + b + {1 \over (1-x)^p} }; \;\;\; \sigma_{\rm c}(x) = \sigma_{\rm s}(x) {\sigma_{\rm 0,c} \over \sigma_{0}} {\alpha(\beta - x)}. 
\ee

\noindent $\sigma_0, \sigma_{\rm 0,c}$ are the strength of the surface and curvature tensions for isospin-symmetric nuclear clusters,  $b$ characterizes the change in the surface and curvature tensions for small deviations from isospin symmetry, while $p$ characterizes them at large isospin-asymmetries. $\alpha, \beta$ are additional parameters that describe the the position and width of the peak of the curvature tension as a function of $x$ \cite{Lorenz1991,LorenzPethick1993,Douchin2000}. Our approach to determining these parameters is outlined in section IV.

The composition of the WS cell is obtained by minimizing the energy of the unit cell with respect to the free variables $n,n_{\rm n},r_{\rm c}$ and $x$. This produces four relations which correspond physically to mechanical, chemical and beta equilibrium of the cell plus the nuclear virial relation
\be \label{eq:virial}
\epsilon_{\rm coul} = 2\epsilon_{\rm surf} + \epsilon_{\rm curv}
\ee

\noindent which expresses the scaling of Coulomb, surface and curvature energy densities under equilibrium with respect to variation of the volume fraction of the charged nuclear component.

To simplify our model, we neglect electron screening and the effect of a neutron skin on the nuclear clusters. Shell effects both in the nuclei and the dripped neutron fluid are not taken into account in this classical model, and the WS approximation is expected to break down at the highest densities in the crust \cite{Chamel2007}. In a later work, using a microscopic calculation, we intend assess the effects of these approximations and missing physics using the results of this paper as a comparison.

%
%

\section{The equation of state of uniform nuclear matter}

Parameters characterizing the EoS of isospin asymmetric nuclear matter around nuclear saturation density $n_{\rm 0}$ are obtained by expanding the EoS in powers of isospin asymmetry $\delta = 1-2x$ and the density parameter $\chi = \frac{n-n_{\rm 0}}{3n_{\rm 0}}$:

\be\label{eq:eos1}
	E(n,x) = E_{\rm 0}(n) + S(n)\delta^2 + ...
\ee
\be\label{eq:eos2}
	E_{\rm 0}(n) = E_{\rm 0} + \half K_0 \chi^2 + ...
\ee
\be\label{eq:eos3}
	S(n) = J + L \chi + \half K_{sym} \chi^{2} + ... 
\ee
\be\label{eq:eos4}
	E_{\rm PNM}(n) \approx E_{\rm 0}(n) + S(n).
\ee
\noindent $E_{\rm 0}(n) = E(n,0.5)$ is the binding energy per nucleon of symmetric nuclear matter (SNM) and $S(n) = \half \partial^2 E(n,x)/\partial \delta^2_{x=0.5}$ is the nuclear symmetry energy. $K_0$ is the incompressibility of SNM at saturation density. $J = S(n_{\rm 0})$, $L = \partial S(n) / \partial \chi |_{n = n_{\rm 0}}$ and $K_{\rm sym}$ are the value of the symmetry energy, its slope and its curvature with respect to density at saturation density. In particular, the pressure of pure neutron matter (PNM) at sub-saturation densities, which plays a large role in determining the equilibrium composition of the crust, can be expressed as
\be \label{eq:eos5}
	P_{\rm PNM } = {n^2 \over 3n_0} [L + (K_0 + K_{\rm sym}) \chi + ...].
\ee  
\noindent to the leading order in $\chi$. 

Our baseline model for the nuclear matter EoS $E(n,\delta)$ is the modified Skyrme-like (MSL) model \cite{MSL01} (see appendix A). The MSL model has the same number of free parameters as the Skyrme model of uniform nuclear matter and they can be analytically related to the properties of nuclear matter at saturation density. This allows for a smooth variation of, e.g., the symmetry energy at saturation $J$ and its slope $L$, while holding fixed the isospin symmetric part of the EoS. 

We note that for all of these EoSs, the symmetry energy is calculated in full without the use of the parabolic approximation (PA) in which equation~(\ref{eq:eos1}) is truncated at second order, thus obtaining the symmetry energy as $S(n) = E_{\rm PNM}(n) - E_0(n)$. The MSL and BD models are parabolic in the potential part of the symmetry energy, but include higher order terms in the kinetic and, in the case of MSL, the effective mass components of the symmetry energy. The PA is sufficiently accurate for small isospin asymmetries since the higher order coefficients are generally predicted to be small compared to second order, but it has been shown \cite{Xu2009,Ducoin2011} that for some NS crustal properties, most notably the crust-core transition density, the PA for the kinetic part of the EoS can lead to divergent predictions compared to the full EoS.

\subsection{Experimental constraints on the symmetry energy}

The magnitude of the symmetry energy at saturation density $J$ is constrained mainly from nuclear mass model fits to experimental data \cite{Myers1966,Moller1995,Pomorski2003,Liu2010}. Care must be taken when interpreting predictions for $J$, since often different definitions for the symmetry energy at saturation are used, for example differing in the use of the parabolic approximation. Moreover, in mass model fits, $J$ depends significantly on what surface symmetry energy is used, e.g. \cite{Danielewicz2009}. We take as a conservative range $25 < J < 35$ MeV.

Constraints on the density dependence of the nuclear symmetry energy have been obtained in analyses of a variety of nuclear experimental data \cite{Li2008,Tsang2009,Tsang2012}. Recent modeling of isospin diffusion in heavy ion collisions involving $^{112}$Sn and $^{124}$Sn extracted constraints of 62$< L <$107 MeV using the IBUU04 transport model \cite{Chen2005,Li2005} and 45$< L <$103 MeV using the ImQMD molecular dynamics model \cite{Tsang2009} respectively. The extraction of the range of $L$ is model dependent; the two ranges quoted here, while they overlap significantly, come from two different transport model analyses. A study of isoscaling in multifragmentation reactions yields $L \sim 66$ MeV \cite{Shetty2007}. Analyses of pygmy dipole resonance (PDR) data gives 27$< L <$60 MeV \cite{Klimkiewicz2007} and the analysis of the surface symmetry energies of nuclei over a wide range of masses gives 75$< L <$115 MeV \cite{Danielewicz2007}. A measurement of neutron skins of a wide mass range of nuclei has led to an estimate of 25$< L <$100 MeV \cite{Centelles2009}. Combining data on the neutron skin thickness of Sn isotopes and isospin diffusion and double n/p ratios in heavy ion collisions, a range of $40 < L < 70$ MeV is obtained \cite{MSL01}, while constraints on the global nucleon optical potential from nucleon-nucleus reactions and single particle energy levels of bound nuclear states leads to a range $30.2 < L < 75.2$ MeV \cite{ChangXu2010}. We take as a conservative range $25 < L < 115$ MeV, noting however that more recent studies tend to favor the lower half of this range \cite{RocaMaza2011, Liu2010}.

To represent the above uncertainties, we will focus on the PNM EoS as this enables us to make contact with microscopic calculations in the following section. For a given model, the SNM EoS ($\delta=0$) and the symmetry energy fixes the PNM EoS. The PNM EoS for the MSL model at a constant value of $J=35$ MeV with $L=25, 70, 115$ MeV is plotted in Fig.~1a.

\begin{figure}[!t]\label{fig:1}
\begin{center}
\includegraphics[width=7.5cm,height=5.5cm]{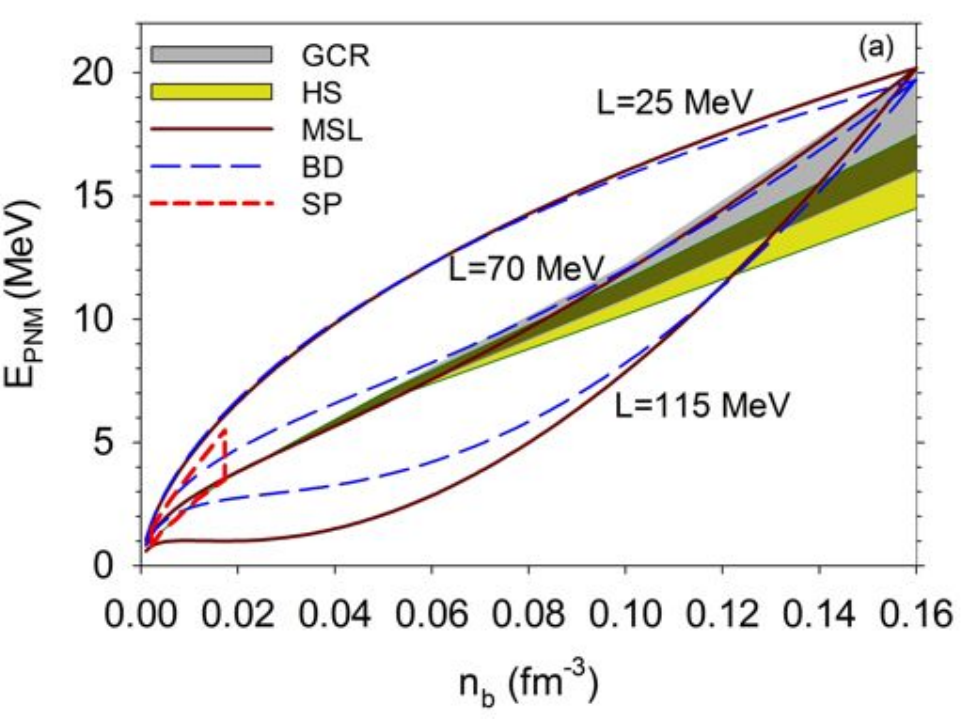}\includegraphics[width=7.5cm,height=5.5cm]{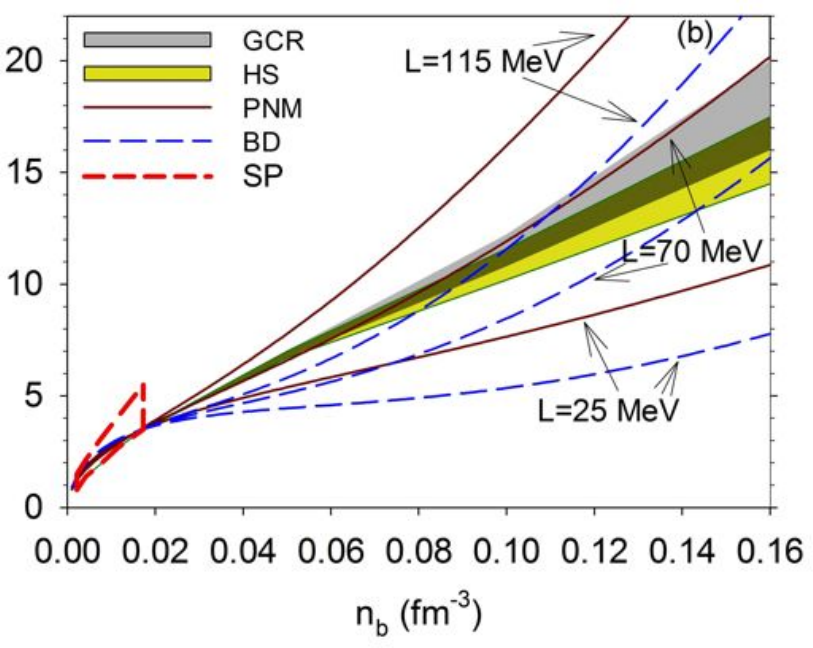}
\caption{(Color online) PNM energy per particle versus baryon number density for the MSL and BD EoSs with $J$=35 MeV (a) and by constraining the MSL and BD EoSs to fit the low-density the PNM EoS from microscopic calculations (b). The gray and dotted bands correspond to the microscopic calculations of Gandolfi et al \cite{Gandolfi2011} (GCR) and Hebeler and Schwenk \cite{Hebeler2010} (HS) taking into account uncertainties in the three-nucleon interaction, while the boxed region at low density comes from the calculations of Schwenk and Pethick \cite{Schwenk2005} (SP).}
\end{center}
\end{figure}

\begin{figure}[!t]\label{fig:2}
\begin{center}
\includegraphics[width=9cm,height=6.25cm]{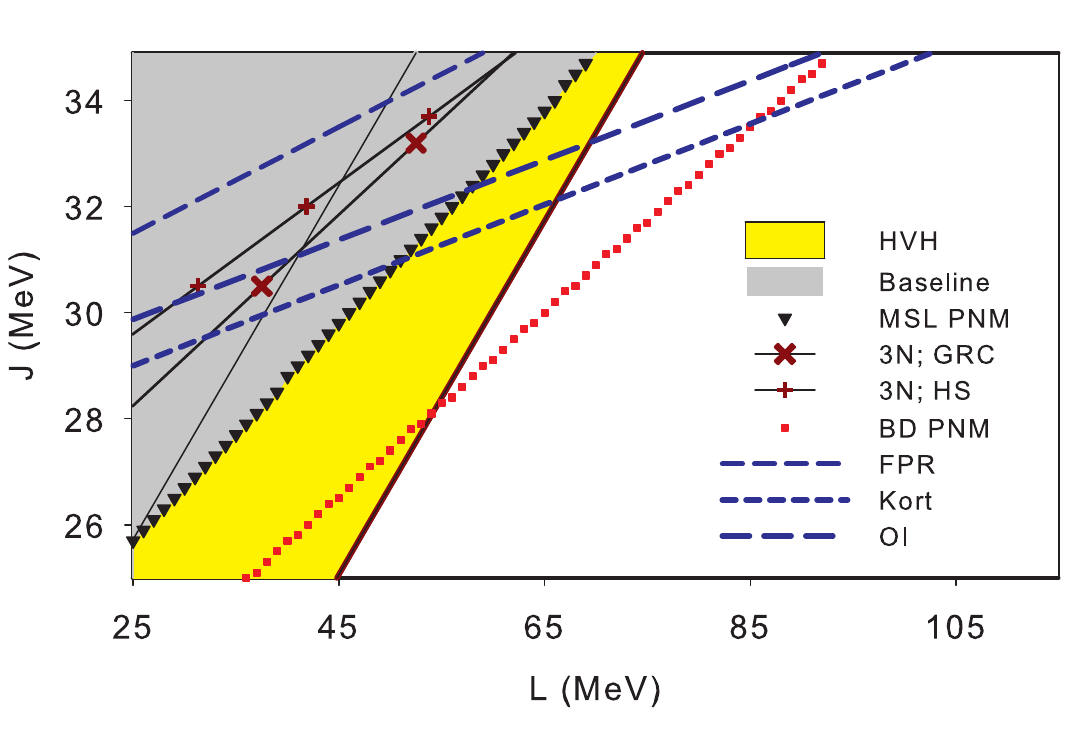}
\caption{(Color online) Correlation between $J$ and $L$ obtained by requiring the MSL and BD EoSs pass through the bounds from low density PNM calculations (triangles and squares respectively); the pink shaded region indicates the region in which the MSL EoS is consistent with the Schwenk and Pethick PNM constraints \cite{Schwenk2005}. The yellow band is obtained by applying the Hugneholtz-van-Hove theorem to optical potential data \cite{ChangXu2010}, and the black lines from PNM calculations at higher densities taking into account uncertainties in the 3-nucleon force (red plusses - chiral perturbation theory \cite{Hebeler2010}; red crosses - QMC methods~ \cite{Gandolfi2011}). The blue lines come from constraints on the saturation properties of asymmetric matter from nuclear mass fits: OI -  \cite{Oyamatsu2007}, FPR - \cite{Farine1978}, Kort - \cite{Kortelainen2010}.}
\end{center}
\end{figure}

\subsection{Theoretical constraints on the pure neutron matter EoS}

Theoretically, the PNM EoS has recently been well constrained at low densities through quantum Monte-Carlo, Green's function Monte-Carlo, chiral effective field theory and variational chain summation techniques \cite{Akmal1998,Carlson2003,Schwenk2005,Gezerlis2008,Hebeler2010}. We show the PNM constraint from Schwenk and Pethick (SP) \cite{Schwenk2005} as the red box at low densities in Fig.~1; results have since converged on the lower bound of that box. Calculations at higher sub-saturation densities introduce uncertainties which are dominated by the theoretical uncertainties in the three-nucleon (3N) force; two recent calculations with estimated theoretical error bars have been performed \cite{Hebeler2010, Gandolfi2011} (labelled HS, GCR hereafter, respectively), and are indicated by the shaded bands in Fig.~1.

The convergence of microscopic calculations using a variety of techniques to the same PNM EoS at low densities, where 3N forces can be neglected, indicates the robustness of these theoretical constraints in that density regime; we shall make use of them to constrain our phenomenological EoS models as follows.

We constrain our MSL EoS to match the PNM EoS region bounded by the microscopic results of SP, HS and GCR below $\approx 0.02$ fm$^{-3}$ - that is the narrow band that runs along the lower edge of the SP results in Fig.~1. For a given value of $L$ we vary $J$ until our phenomenological PNM EoS passes through the constrained region. For each value of $L$, $J$ must be readjusted so that we match the constrained region again. In Fig.~1b, we plot the MSL EoSs obtained following such a procedure. We show results for  $L=25, 70$ and 115 MeV.

One trivial consequence of applying this constraint is that we are no longer free to vary $J$ independently of $L$; we select only those EoSs in $J$-$L$ space which conform to the low density PNM constraints, thus imposing a correlation between the symmetry energy and its slope at saturation density. This correlation is shown in Fig.~2 for the MSL EoS by the black triangles.  Employing the low density PNM constraints, the $J$-$L$ relation obtained is $J$ = 20.5 + 0.207$L$ and $J$ = 18.7 + 0.182$L$ for the MSL EoSs.

The existence of such a correlation is in part due to the variation of the PNM EoS with density; all theoretically derived energy-density functions are simple enough to admit such a correlation. For example, a correlation derived from the low density PNM constraint might not exist if the energy of PNM is not a monotonic function of density in the region up to saturation density. This, however, would imply a region of density in which PNM became stable, a phenomenon that we would expect to have seen indications of in nuclear experiments. Given a smooth, monotonically increasing function of density, a constraint at the low density end will give a correlation between its value at the higher density region, and its slope there.

Although our correlation is simply a result of applying the PNM constraint, such a correlation has been noted in several other works, also plotted in Fig.~2. The PNM calculations can be extended towards saturation density and beyond with the inclusion of uncertain three-body forces; given an empirical value for the symmetry energy at saturation, one then obtains a value for $L$. Correlations in the $J-L$ plane are shown for two such recent studies \cite{Hebeler2010, Gandolfi2011}. The Hugenholtz-Van-Hove theorem predicts a relation between $J$ and $L$ whose uncertainty, in the symmetry potential, can be related to nucleon optical potentials. In the analysis of Xu et al \cite{ChangXu2010}, the yellow shaded region in Fig.~2 is predicted. Nuclear mass fits have been shown to impose a $J-L$ correlation when the symmetry energy parameters are allowed to vary; we show three such relations in Fig.~2 \cite{Farine1978, Oyamatsu2007, Kortelainen2010}. Mass fits tend to give correlations with a smaller slope than our PNM constraints, because they are probing the symmetry energy at higher densities ($\approx 0.1$ fm$^{-3}$ compared to the low density PNM constraints below 0.02 fm$^{-3}$): as one fixes the symmetry energy at densities approaching saturation, the slope of the correlations approaches zero. Studies so far predict correlations that favor the lower-$L$, higher $J$ half of the plane, in which our $J-L$ relation also falls. We show as the dark shaded region those values of $J$ and $L$ obtained with the MSL model consistently with the full range of the earlier SP PNM constraints (the whole of the red dashed boxes in Figs.~1), and we will be highlighting this region of parameter space in our analysis of crust composition to follow. Although we choose the $J-L$ correlation resulting from low density PNM calculations because they are a more direct constraint on uniform nuclear matter, the correlations resulting from nuclear mass fits also fall broadly within the highlighted region of the $J-L$ plane.

\subsection{Higher order symmetry energy parameters}

\begin{figure}[!t]\label{fig:3}
\begin{center}
\includegraphics[width=9cm,height=6.25cm]{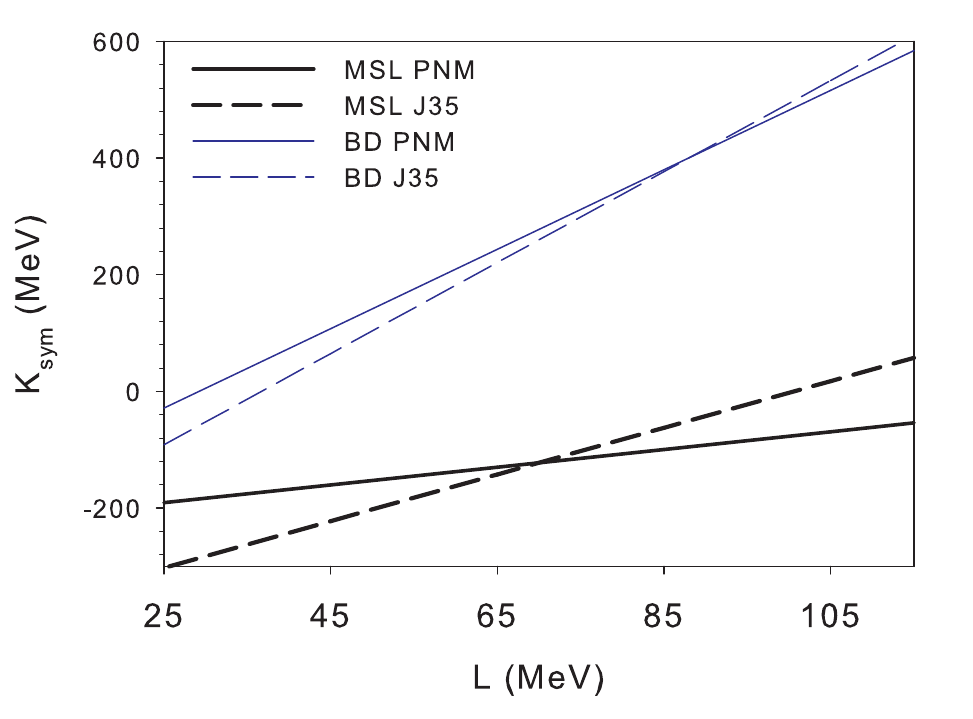}
\caption{(Color online) Correlation between $K_{\rm sym}$ and $L$ obtained from the MSL EoS (black bold lines) and BD EoSs (thin blue lines) for $J$=35 MeV (solid lines) and constrained by low density microscopic calculations of PNM (dashed lines).}
\end{center}
\end{figure}

While $J$ and $L$ characterize the behavior of the symmetry energy around saturation density well, at lower densities $\approx 0.5 n_{\rm s}$ relevant to the properties of NS crusts, additional parameters are necessary to characterize the density dependence of the symmetry energy. Most notably $K_{\rm sym}$, the curvature of the symmetry energy at saturation density, provides information about the lower density dependence. Typically, nuclear models such as Skyrme and Relativistic Mean Field (RMF) models give a relationship between $J$, $L$ and $K_{\rm sym}$ \cite{MSL01,Ducoin2011} such that specifying $J$ and $L$ fixes the value of $K_{\rm sym}$. Relations for $K_{\rm sym}$ versus $L$ for the MSL models with $J=35$ MeV and with $J$ fixed by the PNM constraint are plotted in Fig.~3 (the black bold lines). Skyrme models tend to predict values of $K_{\rm sym}$ and $L$ that follow closely those given by the MSL model \cite{MSL01,Ducoin2011}. However, in order to probe the $K_{\rm sym}$ dependence of our crust EoSs, we shall also use a uniform nuclear matter EoS which predicts a quite different range for $K_{\rm sym}$ for the same values of $J$ and $L$. We use the Bludman and Dover \cite{Bludman1981} mdoel (which we will refer to as BD, see appendix A), which was later modified and used to study finite nuclei and inner crust composition by Oyamatsu and Iida (OI) \cite{Oyamatsu2003, Oyamatsu2007}. The BD model has two fewer free parameters than MSL since it sets the effective nucleon masses to be the same as the bare masses. Like the MSL model, we will examine the cases when $J$ is held fixed and when the BD EoS constrained by the same PNM constraints we use for the MSL model. The $J-L$ correlation obtained from the BD model under the PNM constraint is shown as the red squares in Fig.~2 and is given $J = 18.7 + 0.182 L$. The resulting $K_{\rm sym}$ versus $L$ relations for the BD EoS are shown in Fig.~3, where it is apparent that the model predicts significantly higher values of $K_{\rm sym}$ than the MSL EoS.

The different behavior of $K_{\rm sym}$ in the BD EoS leads to a different density dependence of the PNM EoS at sub-saturation densities important for the behavior of crust properties. We plot the PNM EoS of the BD model alongside the MSL PNM EoS Fig.~1. The differences are most notable in Fig.~1b, where we plot the BD PNM EoS constrained by the same microscopic PNM calculations as the MSL EoS, for the same values of $L$. For a given value of $L$ the MSL and BD EoSs differ significantly, with the BD PNM energy per particle being appreciably lower for most of the density range up to saturation (corresponding to a lower symmetry energy $J$) and its slope shallower; thus the BD EoS will give a lower PNM pressure.

We shall explore the effects of the different $K_{\rm sym}$ behaviors, and consequently the different sub-saturation PNM behaviors, on crust properties in section~6.

%
%

\section{The surface energy}

We now turn our attention to determining the parameters in the expression for the surface energy in Eqs.~7, $\sigma_0, \sigma_{\rm 0,c}, b, p, \alpha, \beta$. These are usually determined from fits to calculations of the interfacial energy between two different phases of semi-infinite nuclear matter (SINM) using, e.g., the Thomas-Fermi or Hartree-Fock methods \cite{Lorenz1991,Douchin2000}, or from fitting liquid drop or droplet models to nuclear mass data \cite{Myers1969,Moller1995,Steiner2008,Danielewicz2003,Steiner2005,Reinhard2006,Danielewicz2009}. The magnitude of the surface and curvature tensions $\sigma_0$, $\sigma_{\rm 0,c}$ for isospin symmetric systems  are relatively well constrained by such methods to be $1.0 \lesssim \sigma_0 \lesssim 1.1$ MeV fm$^{-2}$, $\sigma_{\rm c,0} \approx 0.6$ MeV fm$^{-1}$; in our model we take as our baseline values $\sigma_0 = 1.1$ MeV fm$^{-2}$ and $\sigma_{\rm c,0} = 0.6 $ MeV fm$^{-1}$. We shall also explore the effect of changing $\sigma_0$ to $1.0$ MeV fm$^{-2}$.

Analogously to uniform nuclear matter, the surface and curvature tension may also be expanded about their isospin symmetric values:
\be \sigma_{\rm s}(x) = \sigma_0+ \sigma_{\delta} \delta^2, \ee
\noindent where $\sigma_{\delta} = \half \partial^2 \sigma_{\rm s} / \partial \delta^2 |_{x=0.5}$ is the surface symmetry tension. The parameterization of the surface energy given in Eq. \ref{eqn:surf} then gives an expression for the surface symmetry tension
\be \label{eqn:sdelta}
    \sigma_{\delta} = \sigma_0 {2^p p(p+1) \over 2^{p+1} + b}.
\ee

It is known that the surface symmetry tension (or equivalently surface symmetry energy $S_{\rm s} = 4\pi \sigma_{\delta} (3/4\pi n_0)^{2/3} $) is correlated with the bulk symmetry energy $J$ \cite{Reinhard2006,Steiner2005}. In extracting the surface symmetry energy, two different, thermodynamically consistent formalisms can be employed: the so-called `$\mu_{\rm n}$' and `$\mu_{\alpha}$' approaches which differ in the way the neutron skin is taken into account \cite{Steiner2005}. A roughly linear correlation of $S_s/J$ with $J$  is found from either mass model fits or SINM calculations using a wide range of nuclear models  \cite{Steiner2005}. The correlations differ in slope however, depending on which of the two formalisms is used. The slope of this relation thus neatly encodes the model dependence in the extraction of surface energy parameters. In order to probe this model dependence, we calculate the surface symmetry energy from the bulk symmetry energy using a linear parameterization of $S_s/J$ as a function of $J$ with a slope $c$ which translates to the following relation between surface symmetry tension and $J$:

\begin{figure}[!t]\label{fig:4}
\begin{center}
\includegraphics[width=8cm,height=6cm]{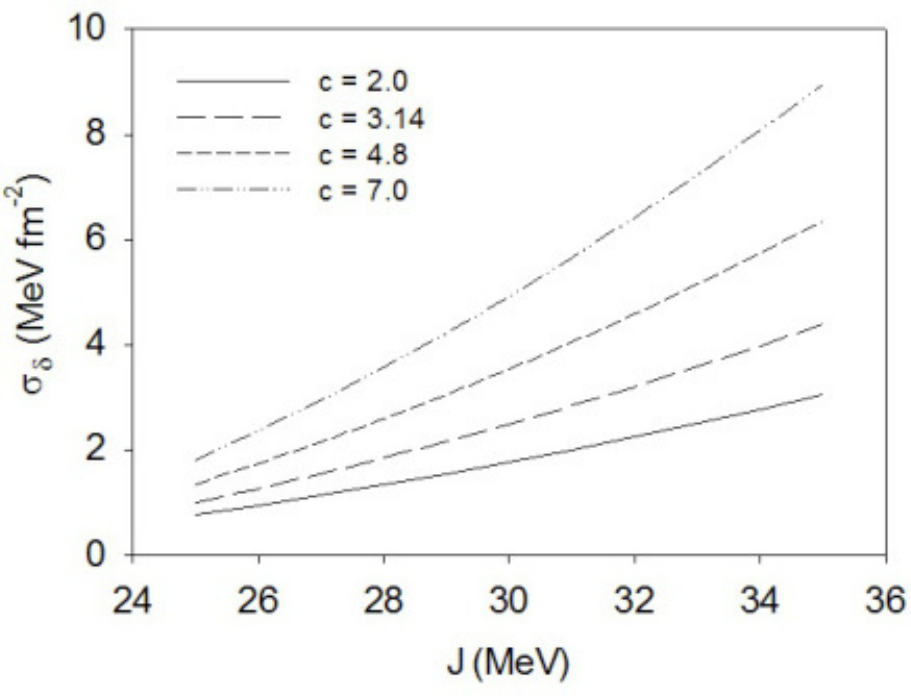}
\caption{Surface symmetry parameter $\sigma_{\delta}$ versus symmetry energy at saturation density $J$ for four values of the slope of the surface-bulk symmetry energy correlation $c$.}
\end{center}
\end{figure}

\be \label{surf_corr} \sigma_{\delta} = { J n_{\rm s}^{2/3} \over (36 \pi)^{1/3}} [ (0.046 \; {\rm MeV}^{-1} c + 0.01\; {\rm MeV}^{-1} ) J - c ]. \ee

\begin{figure}[!t]\label{fig:SurfCurv}
\begin{center}
\includegraphics[width=5.3cm,height=5cm]{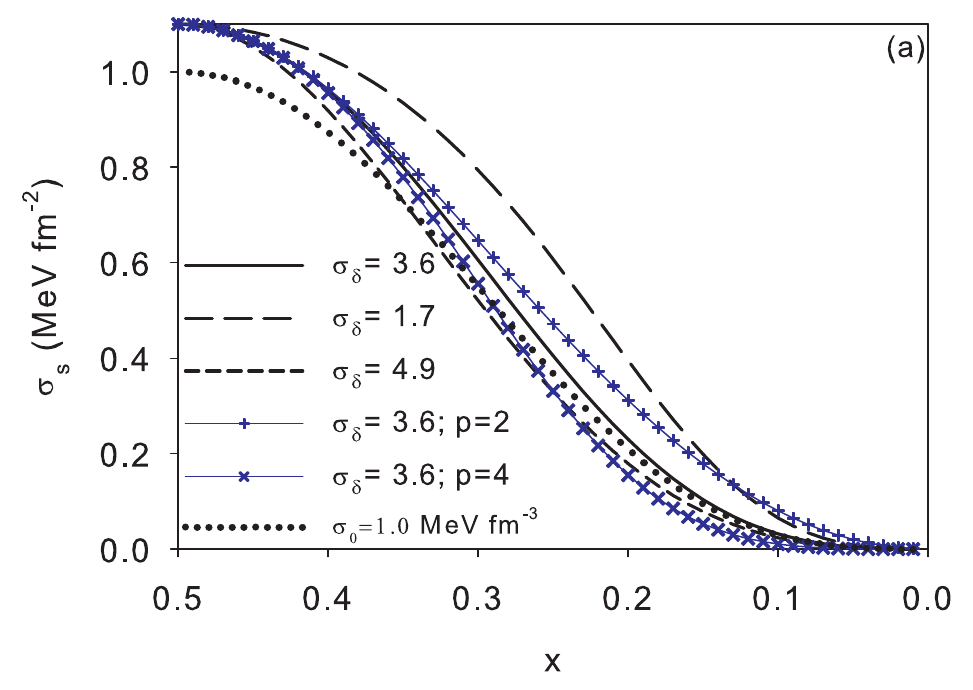}\includegraphics[width=5.3cm,height=5cm]{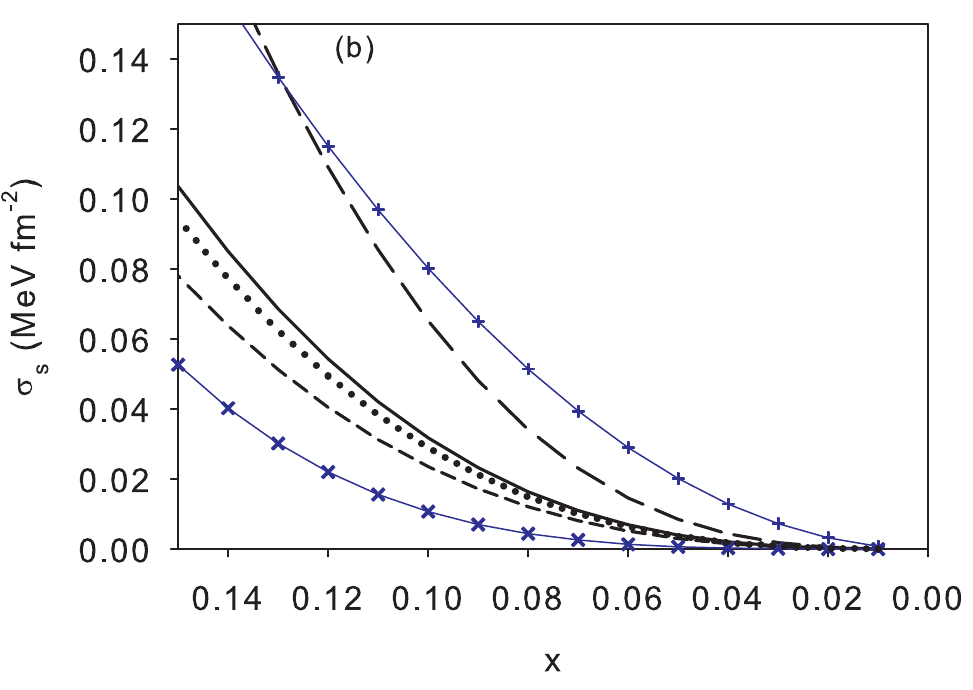}\includegraphics[width=5.3cm,height=5cm]{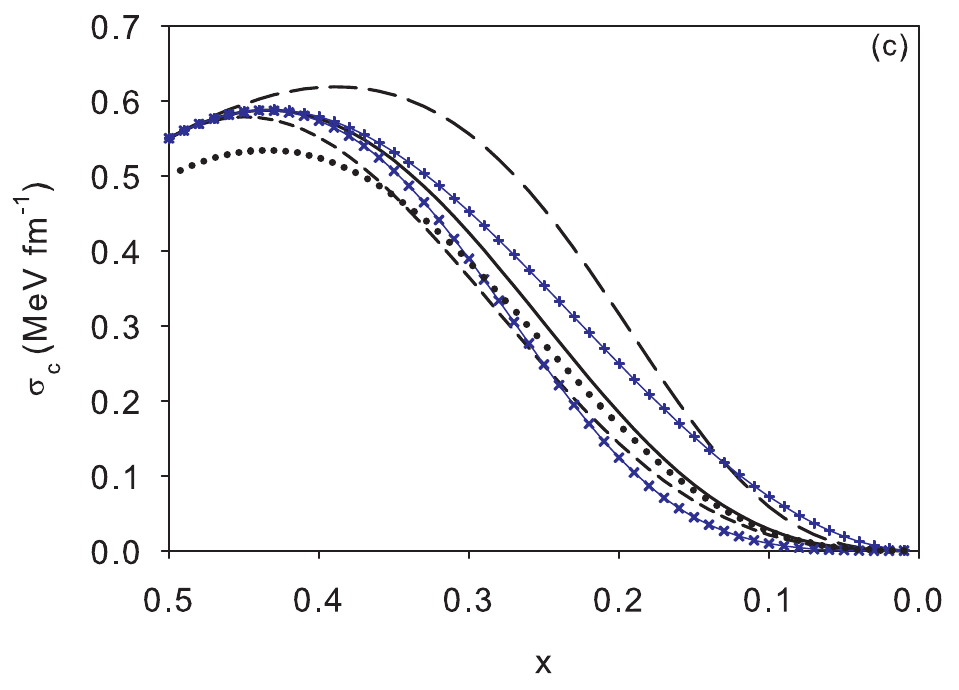}
\caption{(Color online) Surface (a,b) and curvature (c) tensions as a function of proton fraction $x$ for 3 different values of the surface symmetry energy $\sigma_{\delta}$ (1.7,3.6,4.9) corresponding to $J$=30MeV and $c$ = 2,4.8 and 7 respectively, and 2 different parameterizations $p$ of the low proton fraction behavior. The surface tension is shown both over the full range of $x$ (a) and focussing on the low proton fraction region (b).}
\end{center}
\end{figure}

\noindent This approximates well the correlation between surface and volume symmetry energies explored in \cite{Steiner2005}. The free parameter $c$ controls the slope of the correlation and will be varied to explore the model dependence discussed above. Once  $\sigma_{\delta}$ is determined from Eq.~(\ref{surf_corr}), and the parameter $p$ specified in Eq.~(\ref{eqn:sdelta}), the parameter $b$ in the same equation can be calculated. Thus the surface symmetry tension is parameterized by $c$ and $p$ in this work.

The best fits obtained for the  `$\mu_{\rm n}$' and `$\mu_{\alpha}$' approaches  \cite{Steiner2005} are given by $c\approx3.14$ and $c\approx4.8$ respectively in Eq.~(\ref{surf_corr}); the full region of the surface symmetry - bulk symmetry energy plane can be spanned by $c\approx2.0$ to $c\approx7.0$. We will take for our baseline results $c=4.8$. We can then explore the full range of theoretical uncertainty from extracting surface energy parameters by varying $c$ over the full range from $c=2.0$ to $c=7.0$ (see Fig.~4). In Fig.~5 we show the form for the surface and curvature tensions for 3 different values of the surface symmetry energy $\sigma_{\delta}$ corresponding to $J = 30$ MeV, $c =$ 2, 4.8 and 7 by the black solid, long-dashed and short dashed lines respectively. Note that a higher $\sigma_{\delta}$ corresponds to a \emph{lower} (softer) surface energy at a given proton fraction.

Of particular importance for the NS crust is the behavior of the surface energy at low proton fractions, determined in equation~(\ref{eqn:surf}) by the parameter $p$. Fits to plane interface calculations using a variety Skyrme interactions suggest a value of $p\sim3$ \cite{Ravenhall1983.2,Lorenz1991}, a value we take in our baseline calculations. Recent work on neutron drops suggests that the gradient terms in the Skyrme interactions underestimate the surface energy for pure neutron matter \cite{Gandolfi2_2011}, sugesting that perhaps the value of $p$ should be lower (making the surface energy higher at low $x$). We will explore the range $p=$2 - 4. In Fig.~5 we show the form of the surface and symmetry energy for $p$=2 and 4 for  $\sigma_{\delta}$ = 3.6 corresponding to $c=4.8, J=30$ MeV. The middle panel shows a close-up of the low proton fraction region of the surface tension: note that a lower value of $p$ gives a \emph{higher} surface energy in this region.

We note that the strength of the surface energy for symmetric nuclear matter also correlates with the bulk symmetric matter parameters such as $K_0$. However, in this work we allow them to vary independently. As we shall demonstrate, variation of either $\sigma_0$ or $K_0$ within accepted bounds has little effect on the crust EoS compared to the symmetry energy parameters. Finally, we take the remaining curvature parameters $\alpha$, $\beta$ from the results of \cite{Lorenz1991} to be $\alpha=5.5$ and $\beta=1.0$.

%
%

\section{Baseline Results}

\begin{table}[b]
    \begin{center}
        \caption{Baseline parameters} \label{Tab:2}
        \vspace{0.2cm}
        \begin{tabular}{p{2.3cm} p{2.3cm} p{1.8cm} p{2.8cm} p{2.8cm} p{0.8cm} p{0.8cm}} \hline \rule[-8pt]{0pt}{22pt}
        \hspace{0.2cm} $K_0$ (MeV) & $E_0$ (MeV) & $n_0$(fm$^{-3}$)  & $\sigma_0$ (MeV fm$^{-2}$) & $\sigma_{\rm c,0}$(MeV fm$^{-1}$) & $c$ & $p$  \\ \hline
        \hspace{0.2cm}  240 & -16.0 & 0.16 & 1.1 & 0.6 & 4.8 & 3 \\ \hline
        \end{tabular}
    \end{center}
\end{table}

Our CLDM was tested by accurately reproducing the BBP crust composition and EoS using the same nuclear matter model employed in that work. We turn our attention now to the results of our survey of NS crust predictions using the MSL model. Our baseline SNM EoS and surface parameters are summarized in Table~2. We then vary the slope of the symmetry energy $L$ from 25 - 115 MeV under two different constraints:

\begin{itemize}
\item{Constant symmetry energy at saturation density $J$. Varying $L$ for each value of $J$ in the range 25 - 35 MeV generates a new sequence of EoSs. We will label a given sequence by it's value of $J$, e.g. `J35' for $J=35$ MeV.}
\item{$J$ adjusted for each value of $L$ so that our EoS passes through the theoretical low density PNM constraint. Then $J$ will vary with $L$ according to the relations given in Section~3.2. We will label this sequence of crust EoSs `PNM'.}
\end{itemize}

Using the MSL EoS, the region of the $J,L$ parameter space consistent with both the conservative experimental range of $J=25 - 35$ MeV and the theoretical PNM constraint corresponds to the shaded gray region in Fig.~2; the corresponding region for the various crustal properties will also be shown as a shaded region in subsequent plots.

\begin{figure}[!t]\label{fig:6}
\begin{center}
\includegraphics[width=8cm,height=6cm]{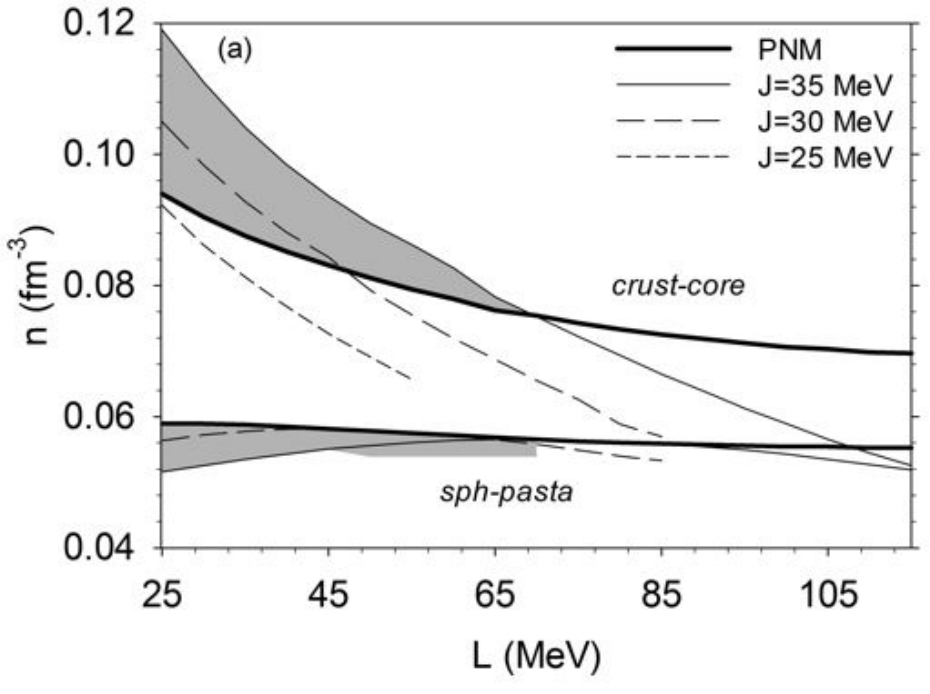}\includegraphics[width=8cm,height=6cm]{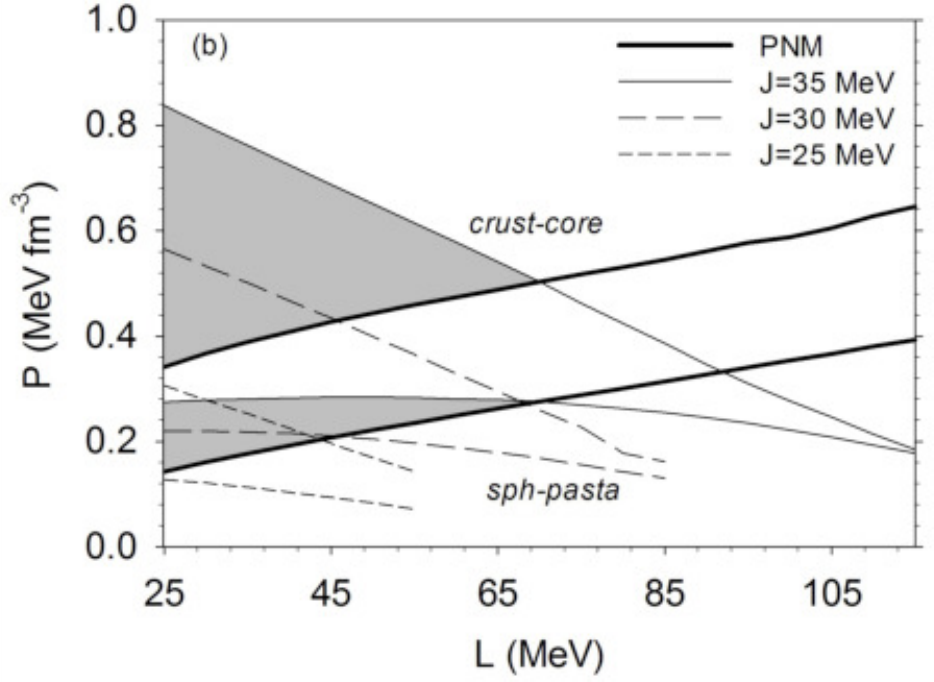}
\caption{Crust-core and spherical nuclei-pasta transition densities (a) and pressures (b) versus $L$. Results are displayed for the 3 constant-J sequences `J35', `J30' and J25' and `PNM' sequence constrained by the low density PNM EoS. The shaded region corresponds to the region consistent with the PNM constraint and $25 < J < 35$ MeV.}
\end{center}
\end{figure}

We will concentrate on the crust-core, and spherical nuclei - pasta transition densities and pressures ($n_{\rm cc}, n_{\rm p}, P_{\rm cc}, P_{\rm p}$ respectively), and on the variation throughout the inner crust of the pressure $P$ and the following four compositional parameters:

\begin{itemize}
\item{$v$, the volume fraction of the charged nuclear component (that is, the component which contains the protons),}
\item{$X_{\rm n}$, the density fraction of free neutrons, $X_{\rm n}= (1-v)n_{\rm n}/n_{\rm b}$,}
\item{$r_{\rm C}$, the radius of the WS cell, and}
\item{$x$, the \emph{local} proton fraction of clustered matter.}
\end{itemize}

\noindent where $n_{\rm n}$ is the local free neutron density and $n_{\rm b}$ the overall baryon density of matter. Any other compositional quantities can be expressed in terms of these four; for example, the mass and charge number of nuclei are given by $A = 4 \pi r_{\rm C}^3 (1 - X_{\rm n})/3$, $Z = xA$ respectively.

In Fig.~6 we show the variation of the crust-core and spherical nuclei-pasta transition densities and pressures over the range of  $L$. Results are plotted for the PNM, J25, J30,  and J35 sequences of crust EoSs.

The crust-core transition density shows the negative correlation with $L$ found in previous works, e.g. \cite{Oyamatsu2007,Xu2009}, for both PNM and constant-J sequences. For the PNM sequence the slope of the correlation is shallower and varies in the range $n_{\rm cc} \approx 0.07 - 0.095$ fm$^{-3}$, a variation of $0.25$ fm$^{-3}$. For the J35 sequence, $n_{\rm cc} \approx 0.05-0.012$ fm$^{-3}$, a variation of $0.07$ fm$^{-3}$. The crust-core transition density also clearly depends on $J$ with a positive correlation, and a variation of $\approx 0.3$ fm$^{-3}$ from $J = 25-35$ MeV is seen independent of $L$. A higher symmetry energy favors a larger proton fraction in the nuclear clusters, lowering their bulk binding energy and thereby making the clustered matter energetically favorable to higher densities. For the `PNM' sequence, $J$ increases with $L$, and so the decrease of the transition density with $L$ is compensated slightly by the increase of the transition density with $J$, making the $n_{\rm cc} -L$ correlation shallower. Indeed, with a sufficiently steep correlation between $J$ and $L$, the correlation can vanish or even be reversed; if we were to take the HVH correlation of Fig.~2, we would see no correlation between $n_{\rm cc}$ and $L$. The shaded region, consistent with our conservative range for $J$ and the PNM constraint encompasses crust-core transition densities of $0.08 - 0.12$ fm$^{-3}$. 

Relative to the crust-core transition density, the spherical nuclei-pasta transition density shows very little variation with $L$ and $J$, tracing out a thin band $n_{\rm p} \approx 0.05-0.06$ fm$^{-3}$. As a consequence of the decreasing crust-core transition density with $L$, the region of density in which pasta structures are predicted to appear also diminishes with $L$, vanishing above $L\approx 90$ MeV for the J30 sequence and above  $L\approx 110$ MeV for the J35 sequence. Although the extent of the pasta region decreases with $L$ for the PNM sequence also, it is always predicted to exist over a significant range of densities over the whole range of $L$ used. 

The transition pressures show different correlations with $L$ depending on whether we choose the PNM constraint (positive correlation) or the constant $J$ constraint (negative correlation). However, it can be seen that transition pressure has a positive correlation with $J$; for the PNM constraint $J$ increases with $L$, and so overall the correlation is determined by the competition between increasing $P_{\rm cc}$ with $J$ and decreasing $P_{\rm cc}$ with $L$. The spherical-pasta transition pressure $P_{\rm p}$ also increases with $L$ for the PNM constraint, whereas the variation with $L$ at constant $J$ varies only weakly with $L$. The crust-core and spherical-pasta transition pressures consistent with the PNM constraint and $25<J<35$ MeV both vary by more than a factor of 2, $P_{\rm cc} \approx 0.35 - 0.85$ MeV fm$^{-3}$, $P_{\rm p} \approx 0.15-0.3$ MeV fm$^{-3}$.

Note that the shaded region is bounded by the PNM and J35 sequences for $L < 70$ MeV; these two sequences will be taken as the bounding regions in what follows.

\begin{figure}[!t]\label{fig:7}
\begin{center}
\includegraphics[width=7.5cm,height=4.65cm]{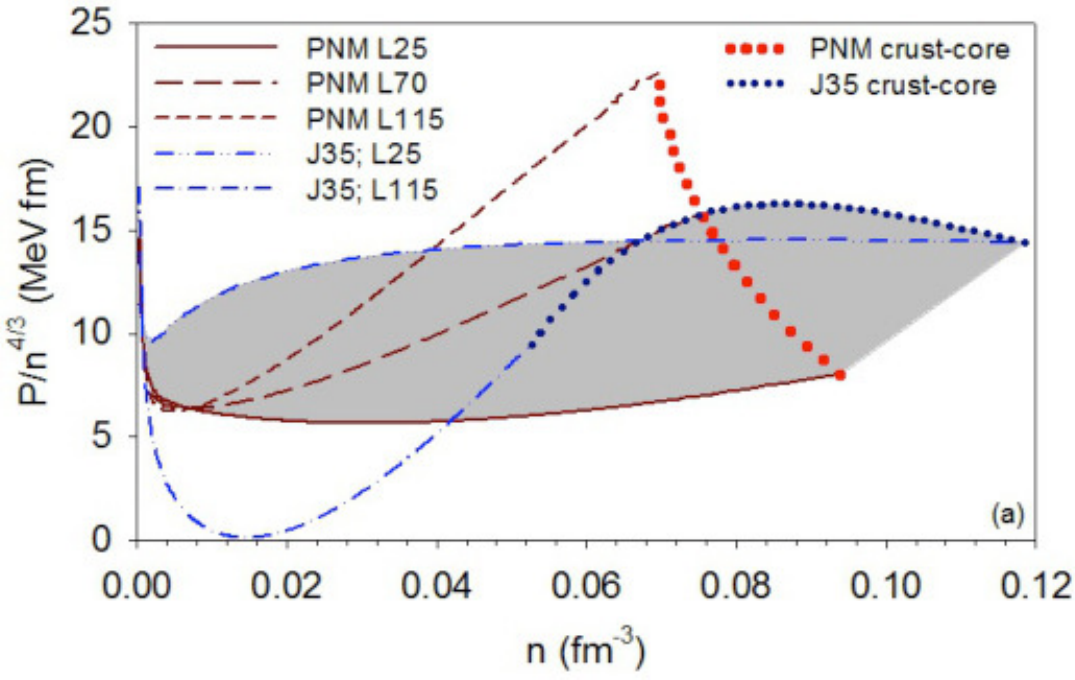}\includegraphics[width=7.5cm,height=4.65cm]{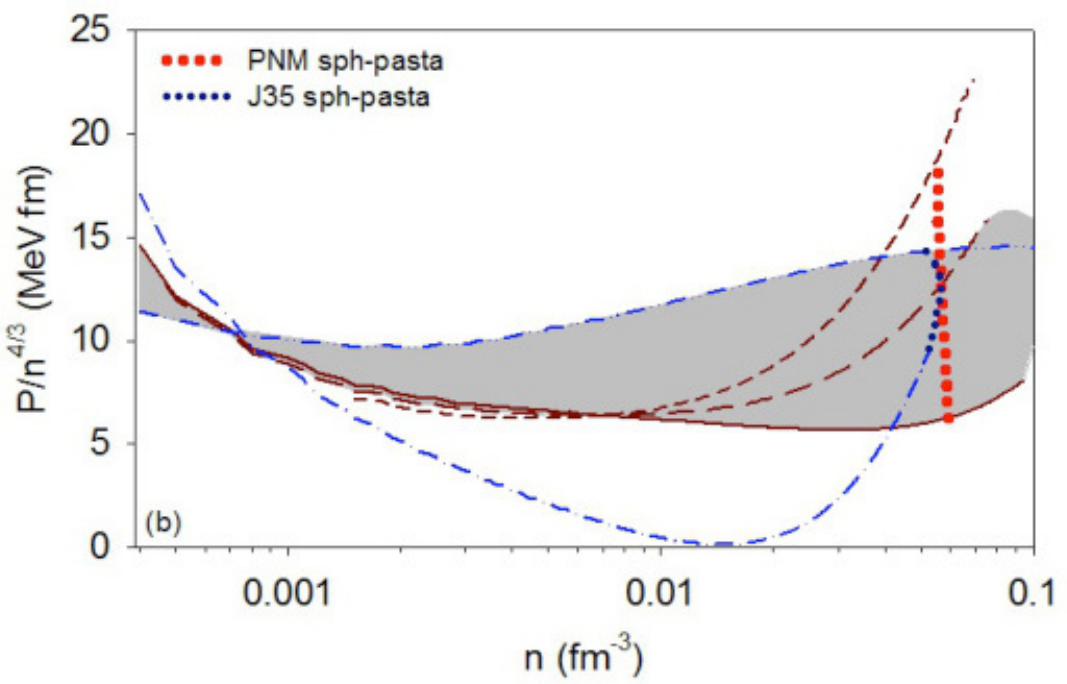}
\caption{(Color online) Pressure of matter scaled by $n^{4/3}$ versus baryon number density in the inner crust. Results are displayed for the $L=25, 70$ and 115 MeV members of the `J35' and `PNM' sequences. Results are displayed over the whole density range of the inner crust (a) and the lower density region up to the transition to pasta (b). The loci of the transition densities over the range $L = 25 - 115$ MeV are displayed for the `PNM' sequence (larger black circles) and the `J35' sequence (smaller blue circles). The shaded region has the same meaning as in Fig.~6.}
\end{center}
\end{figure}

In Figs~7-9 we display the ranges of compositional parameters predicted in the baseline model. Results are shown for the PNM and J35 sequences for $L = 25,70$ and $115$ MeV (note that the two sequences coincide at $L = 70$ MeV). The left panels show the full density range on a linear scale to highlight the density range of the pasta phases. The loci of the crust-core transition densities are shown for both PNM (larger black circles) and J35 (smaller blue circles) sequences. The right panels show the range of densities up to the transition to pasta. The loci of the spherical-pasta transition densities are shown for both PNM (larger black circles) and J35 (smaller blue circles) sequences.

\begin{figure}[!t]\label{fig:8}
\begin{center}
\includegraphics[width=7.5cm,height=4.65cm]{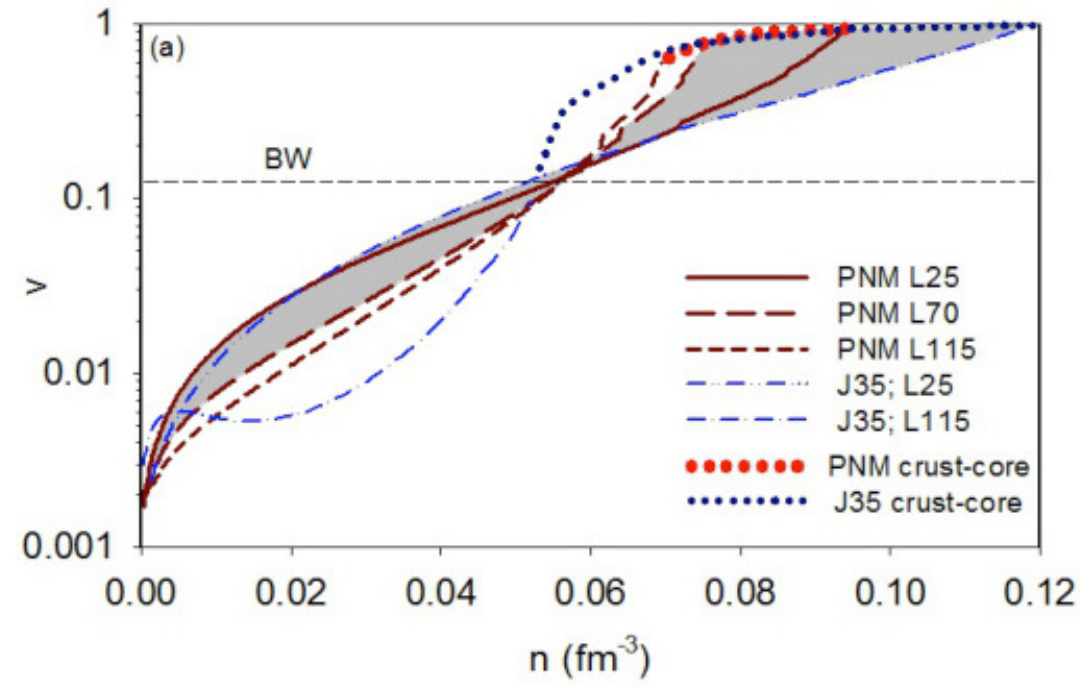}\includegraphics[width=7.5cm,height=4.65cm]{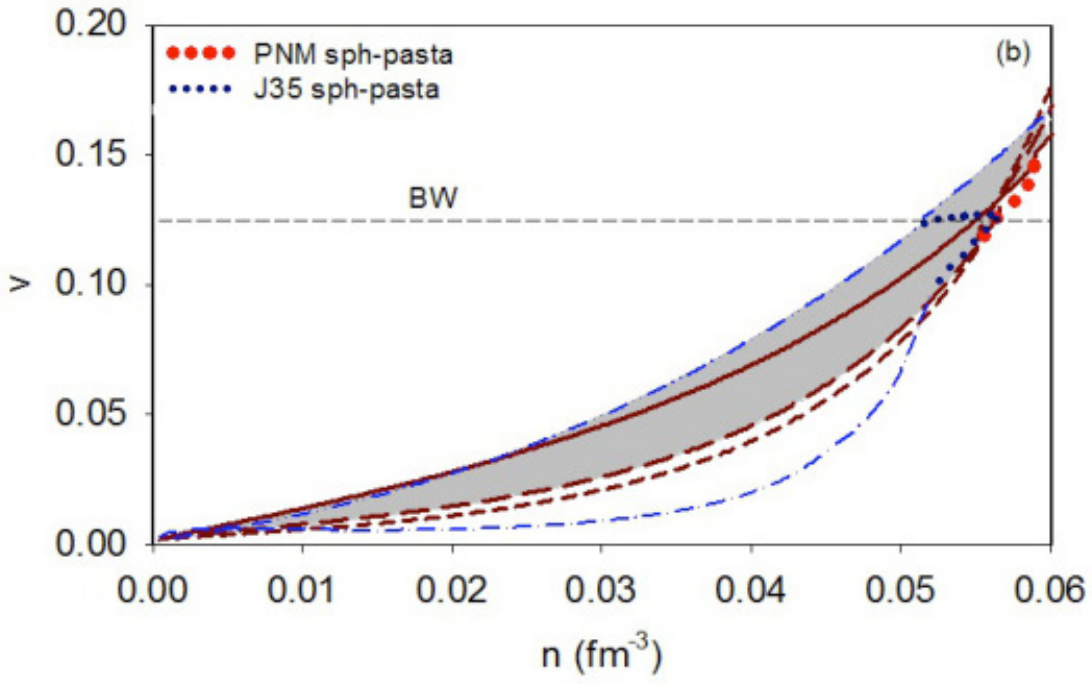}
\includegraphics[width=7.5cm,height=4.65cm]{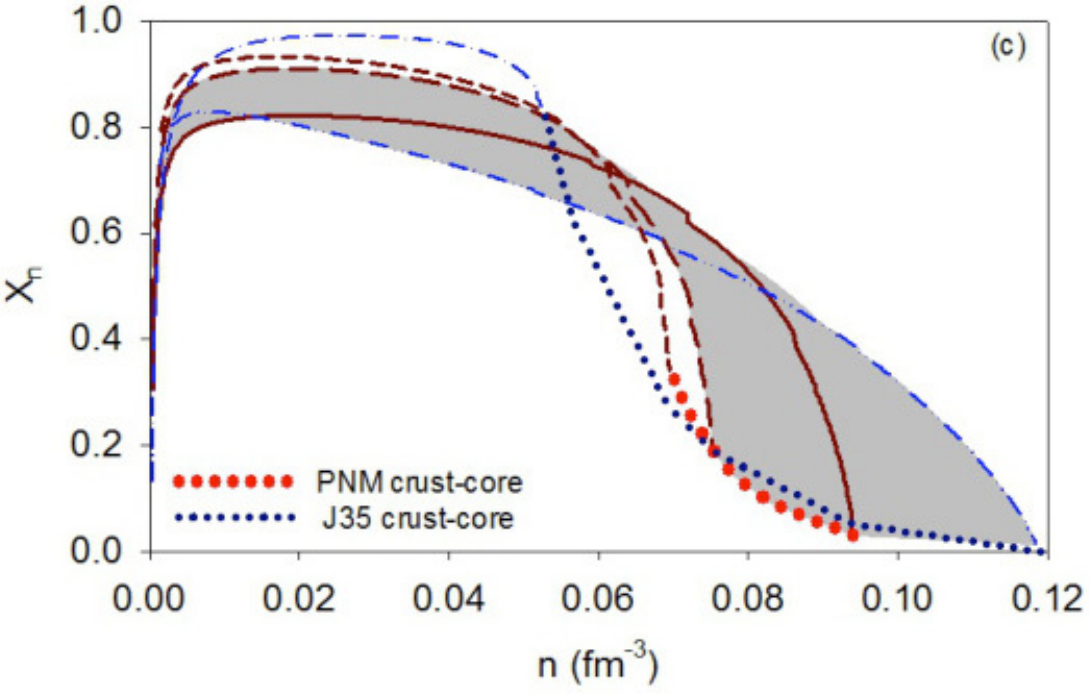}\includegraphics[width=7.5cm,height=4.65cm]{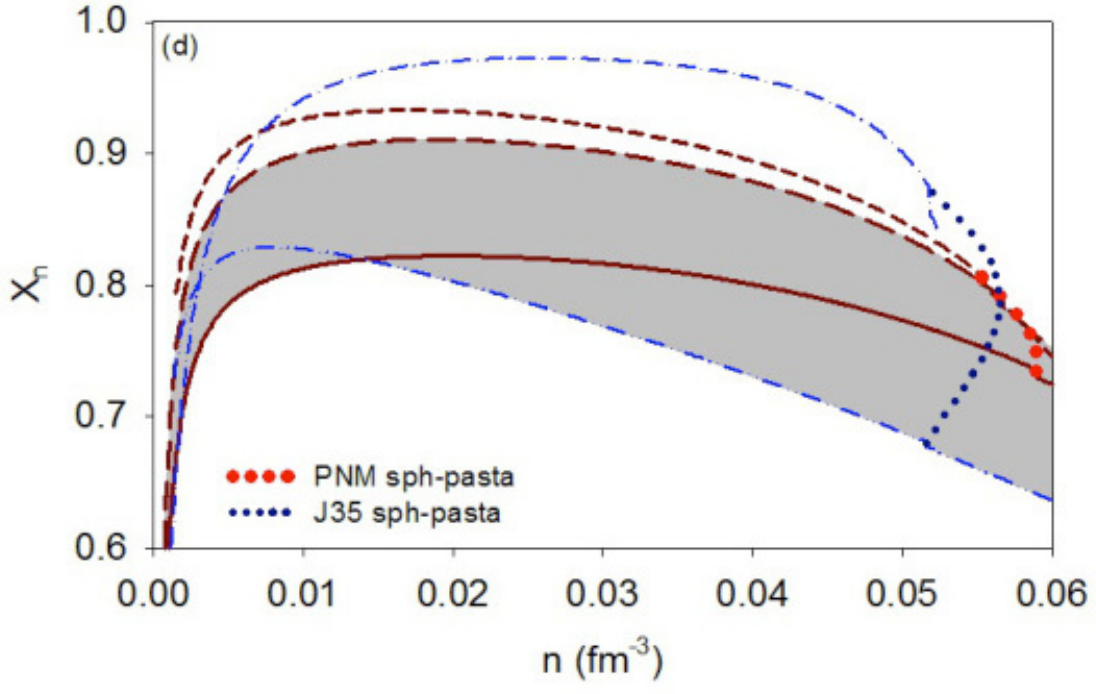}
\caption{(Color online) The volume fraction $v$ occupied by charged nuclear clusters (a,b), and density fraction $X_{\rm n}$ of free neutrons (c,d), versus baryon number density in the crust. The same selection of EoSs from Fig.~7 is used, and the transition loci are displayed in the same way. The shaded region has the same meaning as in Fig.~6.  Results are displayed over the whole density range of the inner crust (a,c) and the lower density region up to the transition to pasta (b,d).}
\end{center}
\end{figure}

Fig.~7 shows the total pressure of matter throughout the crust, scaled to $n^{4/3}$; a free, relativistic neutron gas would result in a constant. Most models, throughout the middle regions of the inner crust, predict a roughly constant trend too. The main deviation comes from the $L=115$ MeV member of the J35 sequence, which predicts a vanishing slope for the PNM energy per particle versus density, and hence a PNM pressure which approaches zero. As expected, since the pressure comes to be dominated by the free neutrons in the crust, a higher value of $L$ leads to a higher pressure throughout most of the crust.

Fig.~8 shows the volume fraction of clustered matter and the density fraction $X_{\rm n}$ of free neutrons. The volume fraction reaches $> 0.8$ at the crust-core transition for the shaded region;  this emphasizes the fact that the WS approximation is certainly not valid at the highest crustal densities. For the PNM constraint, the crust-core transition fraction weakly depends on $L$, ranging from 0.8 ($L$=115 MeV) to close to 1.0  ($L$=25 MeV). For $J=35$ MeV, the crust-core transition fraction varies much more widely, from close to 1  ($L$=25 MeV) down to 0.1  ($L$=115 MeV). The spherical-pasta transition fractions vary over a much smaller range from $\approx$ 0.11 - 0.14 for PNM low and $\approx$ 0.08 - 0.12 for $J$ = 35 MeV. A useful comparison is with the simple estimate from the Bohr-Wheeler fission instability criterion (see, e.g. \cite{PethickRev1995}) of 0.125, a value which closely matches the range predicted by our baseline model. This value is indicated on Fig.~8 by the horizontal line. Below that density, the fraction varies by up to 0.05 from $L$=115 MeV (lower values) to $L=25$ MeV (higher fractions), an easily understood correlation: higher $L$ implies higher neutron matter pressure at saturation density and below, resulting in the pressure equilibrium condition favoring smaller nuclei.

\begin{figure}[!t]\label{fig:9}
\begin{center}
\includegraphics[width=7.5cm,height=4.65cm]{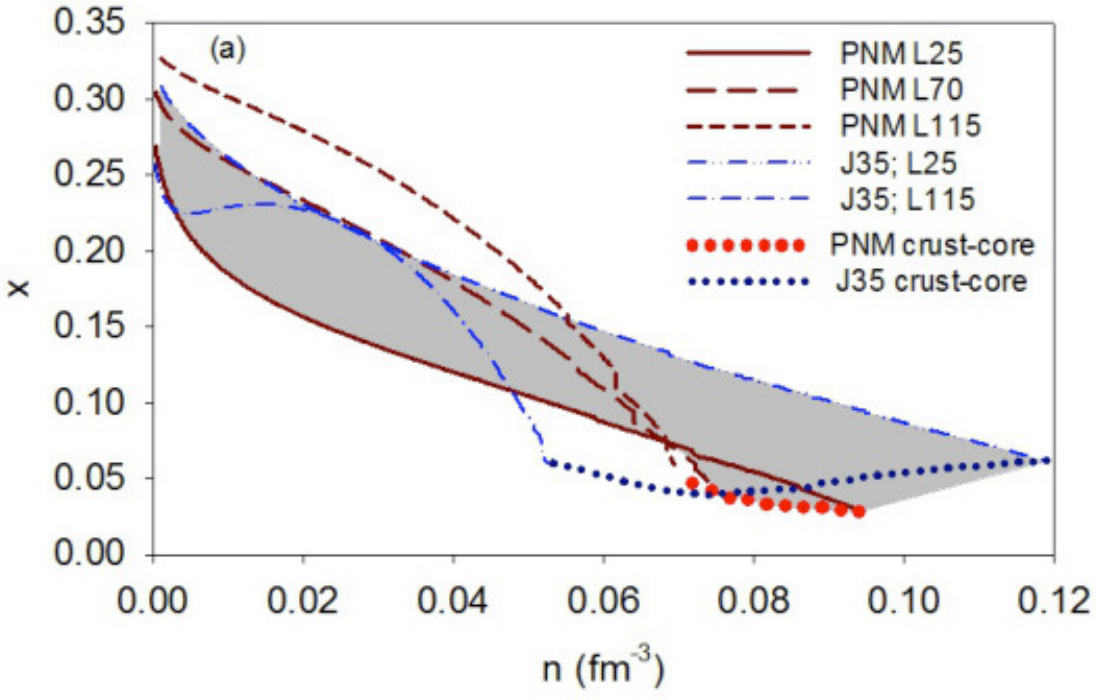}\includegraphics[width=7.5cm,height=4.65cm]{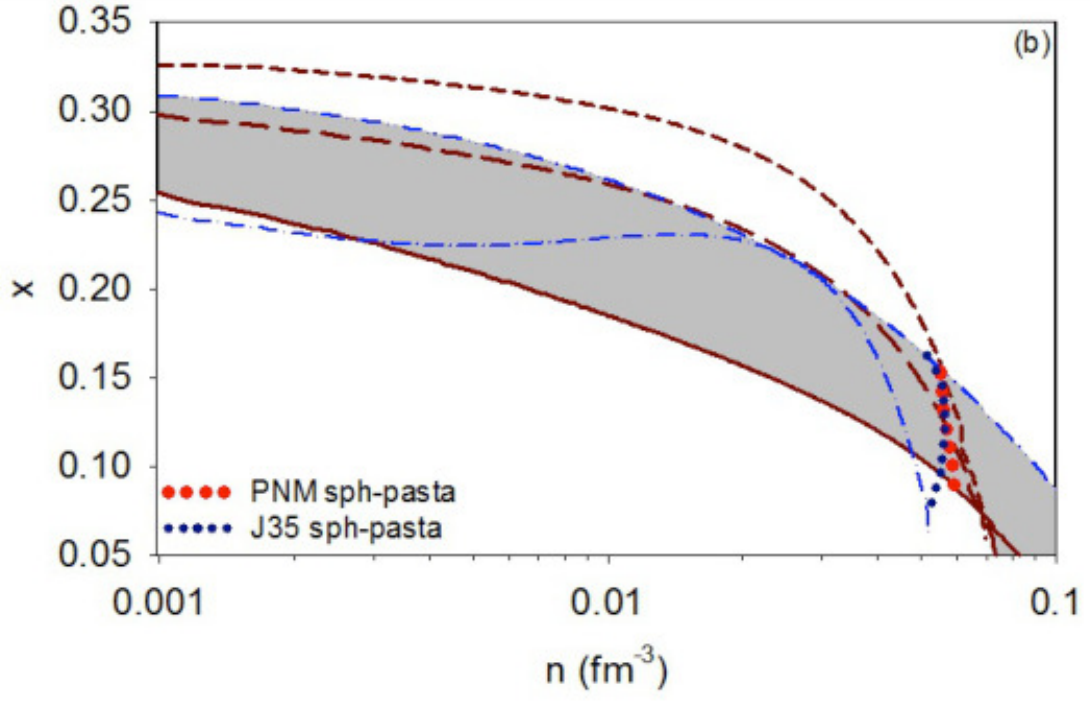}
\includegraphics[width=7.5cm,height=4.65cm]{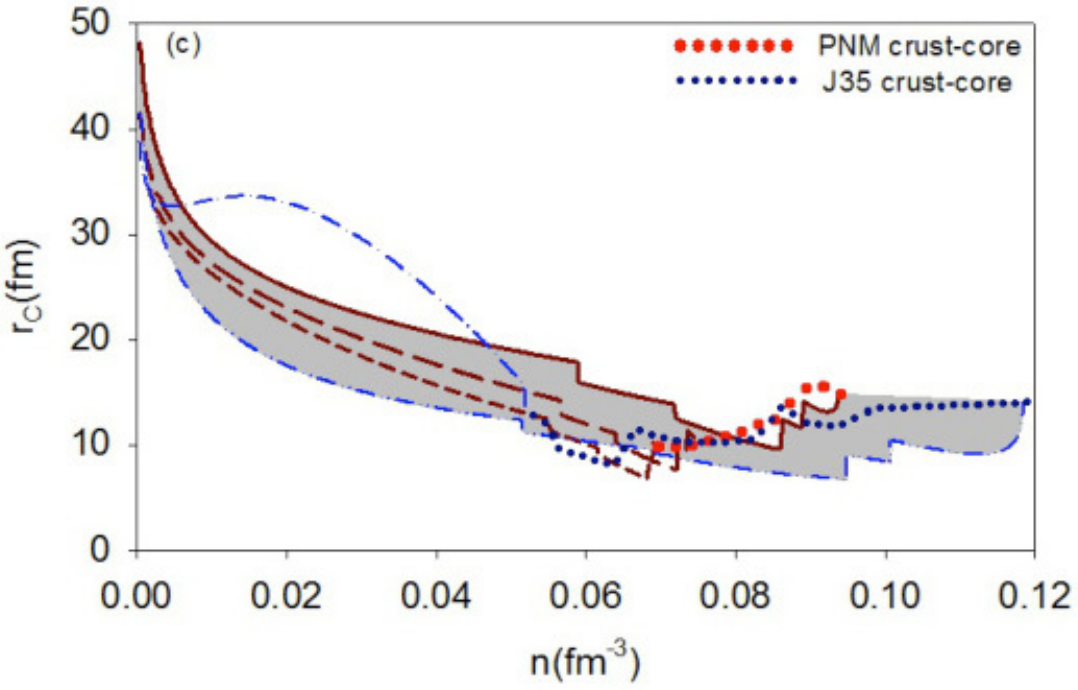}\includegraphics[width=7.5cm,height=4.65cm]{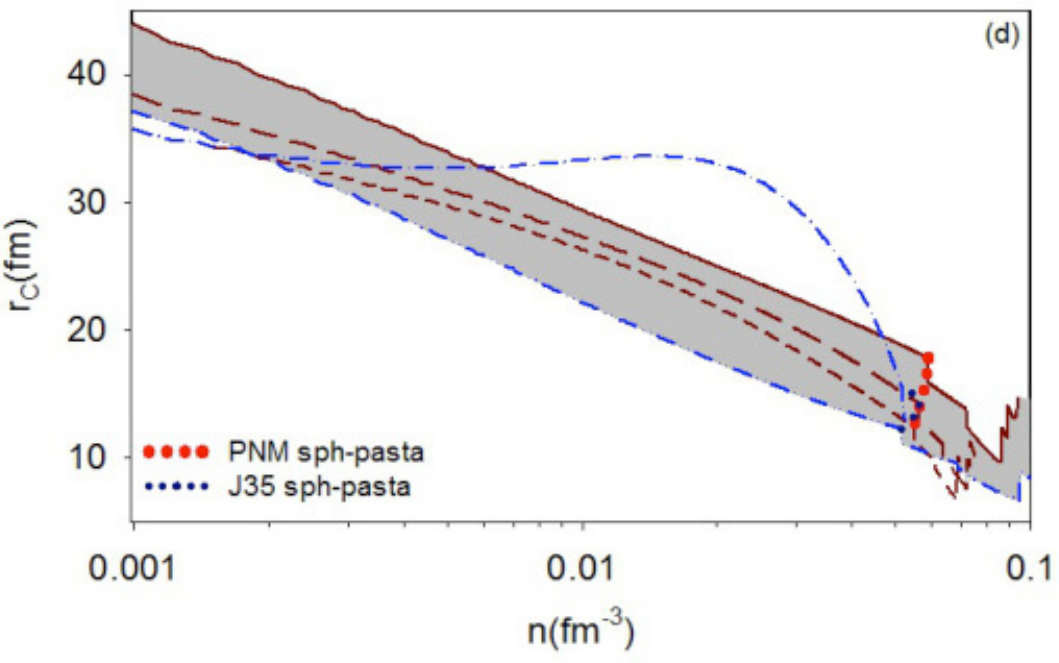}
\caption{(Color online) Local proton fraction of clusters $x$ (a,b) and WS cell size $r_{\rm C}$ (c,d) versus baryon number density in the crust. The same selection of EoSs from Fig.~7 is used, and the transition loci are displayed in the same way. The shaded region has the same meaning as in Fig.~6. Results are displayed over the whole density range of the inner crust (a,c) and the lower density region up to the transition to pasta (b,d).}
\end{center}
\end{figure}

Mirroring the volume fraction at the crust-core transition, $X_{\rm n}$ varies from close to 0.0 up to 0.2. The fraction rapidly converges to zero at low densities as the density of free neutrons vanishes. $X_{\rm n}$ peaks around 0.01 - 0.03 fm$^{-3}$ with values between 0.8 - 0.96, with higher $L$ being correlated with higher $X_{\rm n}$. The transition to pasta occurs close to the peak value of $X_{\rm n}$ for all $L$ and $J$. 

Fig.~9 shows the proton fraction of clustered matter $x$ and the cell size $r_{\rm C}$. $x$ generally decreases with increasing density over the whole density range. For $J=35$ MeV, the crust-core transition fraction varies by about 0.03; from $L=25$ MeV, $x_{\rm cc}$ decreases from 0.06 down to 0.03 at $L=70$ MeV, then increasing back up to 0.06 at $L=115$ MeV; for PNM the variation is similar, starting at $x_{\rm cc}$=0.025 for  $L=25$ MeV and increasing with $L$ up to 0.06 at $L=115$ MeV. In the lower density region the uncertainty in $x$ remains approximately constant at around 0.05. At a given density, higher $J$ and higher $L$ correlate with higher $x$. Higher $J$ favors a higher proton fraction; higher $L$ favors smaller, denser nuclei and nuclear clusters; as the symmetry energy increases with density, denser nuclei will also tend to favor higher $x$.

$r_{\rm C}$ varies smoothly up to the transition to pasta, and then proceeds through a series of discontinuous jumps as matter transitions through the various nuclear shapes. In practice, these jumps may be smoothed out by the existence of intermediate shapes not considered in this work \cite{Nakazato2009}. The negative correlation of $r_{\rm C}$ with $L$ is again explained by the pressure equilibrium condition requiring smaller nuclei for higher values of $L$. In the right panel, it can be seen that the uncertainty in $r_{\rm C}$ due to the uncertainty in symmetry energy behavior at saturation density remains roughly constant over the inner crust at $\approx$ 8 fm.

%
%

\section{Variation of remaining CLDM parameters around baseline values}

Having established how the crustal parameters behave over the range of $L$ and $J$ for our baseline parameter set, we now examine how those results depend on the remaining model parameters describing the surface energy and symmetric nuclear matter. We shall also examine the effect of a different range of $K_{\rm sym}$ by comparing the baseline results to a those obtained with the BD EoS, and we shall also compare our results to two widely used CLDM crust models from the literature (DH \cite{Douchin2001} and BBP \cite{BBP1971}). We shall use our PNM baseline sequence as the reference set of results from the previous section. 

For the surface energy parameters, we shall consider the following variations: (i) surface symmetry - volume symmetry slope parameter $c$ = 2.0 (7.0) corresponding to higher (lower) surface energies at proton fractions close to 0.5 respectively, (ii) $p$ = 2.0 (4.0) corresponding to higher (lower) surface energies at low proton fractions $x \to 0$ respectively (see Fig.~5), and (iii) $\sigma_0$ = 1.0 MeV fm$^{-3}$. For SNM, we consider variations of (i) $K_0$ = 220 - 260 MeV and (ii) $n_0$ = 0.14 - 0.17 fm$^{-3}$.

Since the saturation properties of nuclear matter, coupled with fixing the symmetry energy $S(n)$ at some density $n$ are insufficient to uniquely constrain the density dependence of the PNM EoS at sub-saturation densities, we shall examine the effect of using a different PNM EoS by using the BD model (see section~3). As Fig.~1 illustrates, the MSL and BD EoSs, constrained by the same low density PNM EoS, differ significantly at intermediate densities because of their different symmetry energy curvatures $K_{\rm sym}$. The BD EoS gives higher values of $K_{\rm sym}$ leading to a lower energy and pressure of PNM for a given value of $L$ along the PNM sequence.

Of the two other CLDMs we compare to, BBP is now known to use a surface energy and incompressibility at odds with recent experimental results and thus we expect it to give divergent results compared to our model. DH constructed their CLDM using the SLy4 Skyrme parameterization of the nuclear matter EoS, and determined surface and curvature parameters consistently. They also include the effects of a neutron skin.

\subsection{Transition densities and pressures}

Fig.~10 displays the transition densities and pressures for variations in the surface parameters (left two plots), and symmetric EoS parameters (right two plots). We also include the results for the BD model on the right two plots. In what follow we shall denote the crust-core transition densities and pressures $n_{\rm cc}$ and $P_{\rm cc}$ respectively and the spherical nuclei-pasta transition densities and pressures $n_{\rm p}$ and $P_{\rm p}$ respectively.

\subsubsection{Variation of surface parameters}

\begin{figure}[!t]\label{fig:10}
\begin{center}
\includegraphics[width=8cm,height=6cm]{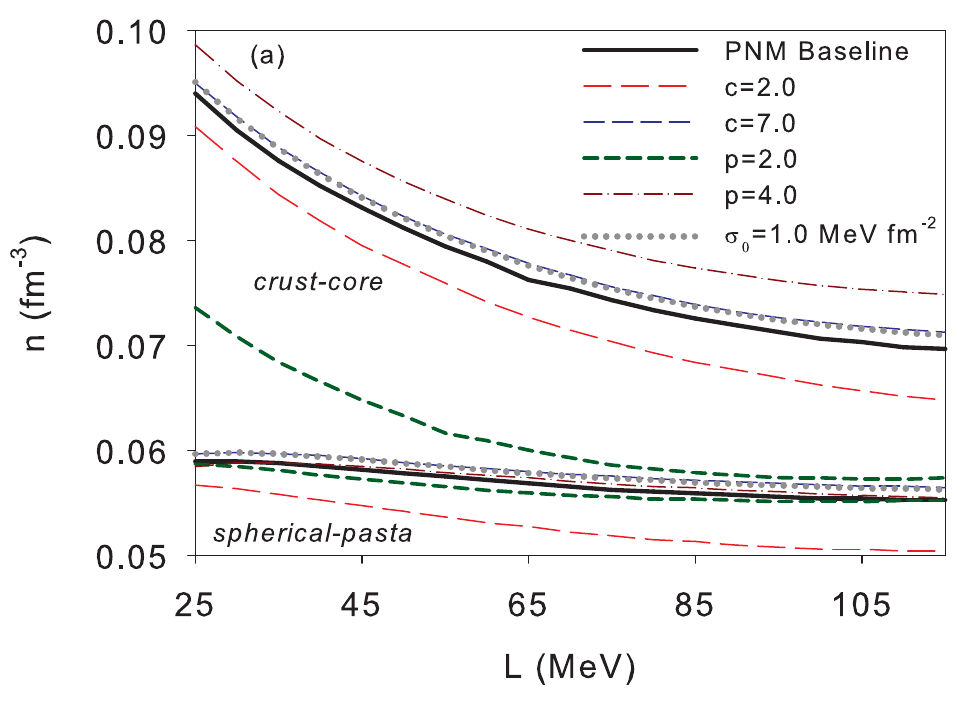}\includegraphics[width=8cm,height=6cm]{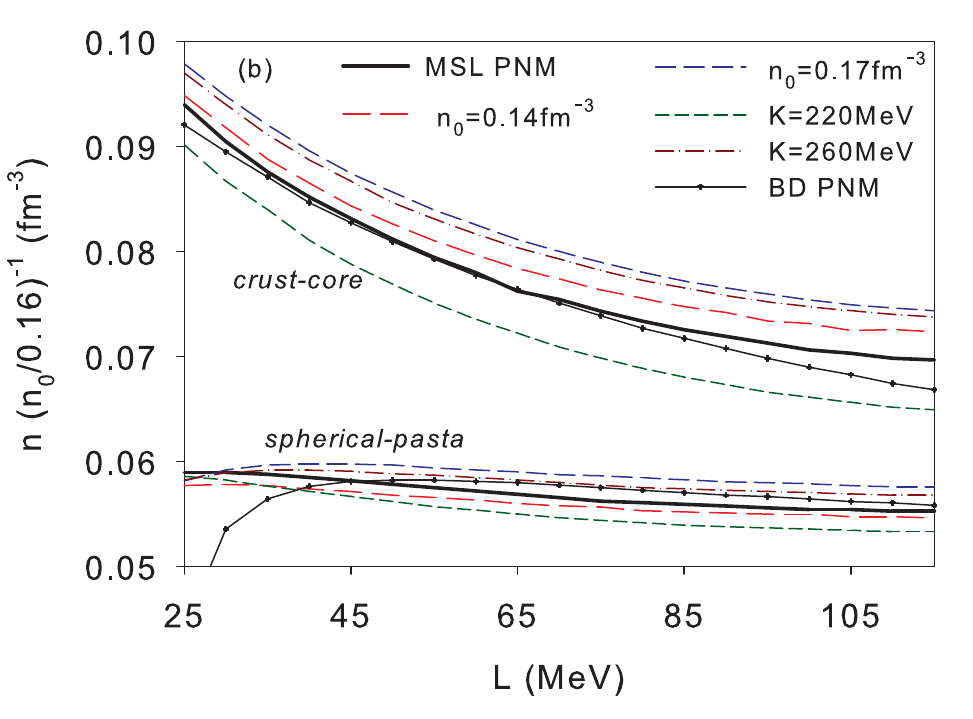}
\includegraphics[width=8cm,height=6cm]{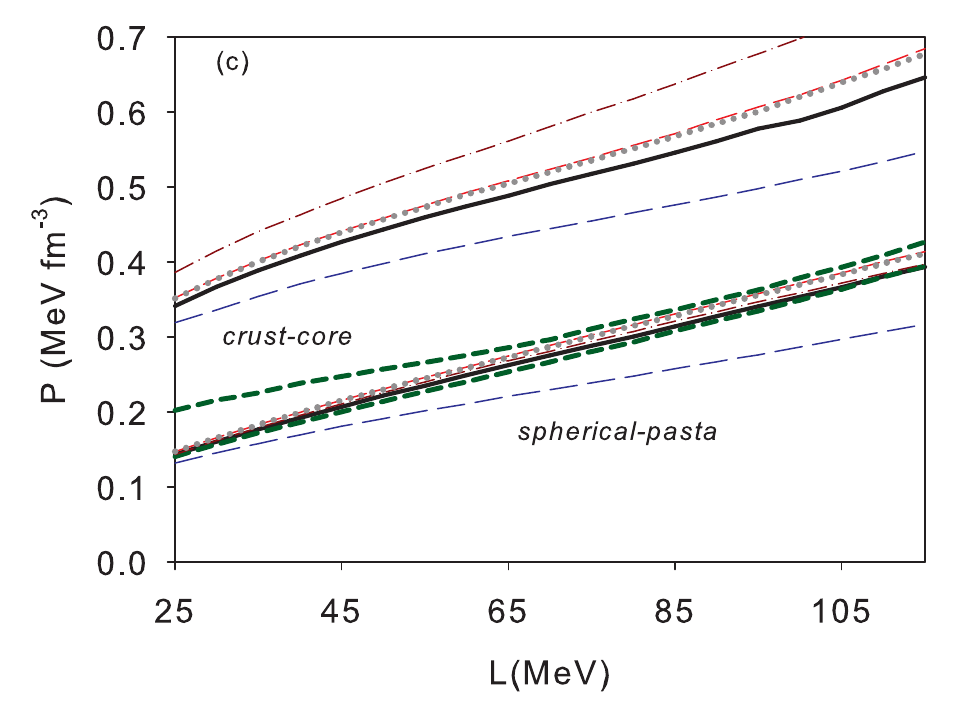}\includegraphics[width=8cm,height=6cm]{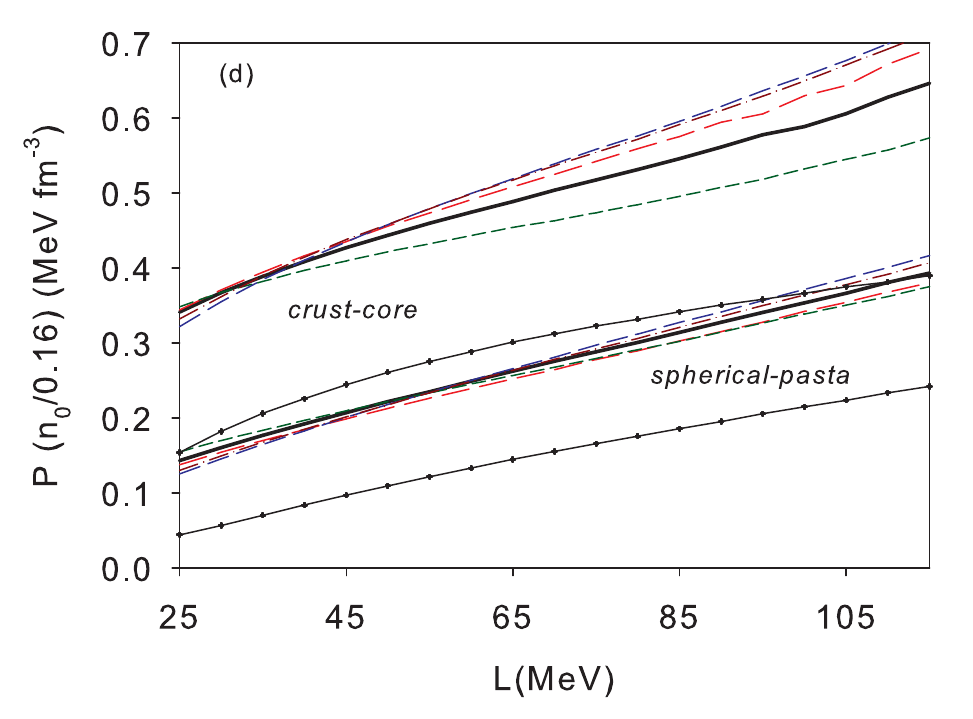}
\caption{(Color online) Crust-core and spherical nuclei-pasta transition densities (a,b) and pressures (c,d) versus $L$ for different parameterizations of the surface energy (a,c) and variations in the symmetric nuclear matter EoS (b,d) compared to the baseline PNM model (thick solid lines in all figures). A lower $c (p)$ corresponds to a higher surface energy at high (low) proton fractions. A stiff surface energy at low proton fractions ($p$=2) results in a notably lower crust-core transition density and pressure, highlighted by the thick, short dashed line. Plots (b,d) also include the PNM model of a different functional form of the nuclear EoS (BD) which predicts lower energy per particle and pressure for PNM at sub-saturation densities than the MSL EoS with the same values of $J$ and $L$ (thin dotted lines). }
\end{center}
\end{figure}

A higher surface and curvature energy in asymmetric nuclear clusters (lower $c$ or $p$) will raise the total energy density of clustered matter compared to uniform matter at the same density, and thus tend to favor a lower $n_{\rm cc}$ for a given value of $L$. For $P_{\rm cc}$, the situation is a little more complicated: for a given density, a higher surface and curvature energy gives a higher pressure, but if the transition density is sufficiently lowered, the transition pressure will also be lowered overall. For the transition to pasta, there is a competition between surface and curvature energies: spherical nuclei have a larger curvature energy than the pasta shapes, so a higher curvature energy will lower $n_{\rm p}$; however, the surface energy will also rise, and since the cylindrical pasta nuclei tend to have a larger surface area compared to the cell surface than spherical nuclei, this will tend to increase $n_{\rm p}$. We see these trends manifest in plots a and c. The $c=2.0$, $p=2.0$ sequences have lower $n_{\rm cc}$ and $n_{\rm p}$ compared to the baseline ($c=4.8$, $p=3.0$) results, while the $c=7.0$, $p=4.0$ sequences have elevated $n_{\rm cc}$ and $n_{\rm p}$.

The effect on $n_{\rm cc}$ is relatively small when $c$ is varied, amounting to $\approx \pm 0.005$ fm$^{-3}$ around the baseline results; for $P_{\rm cc}$ the effect is small for lower $L$, rising to $\approx \pm 0.1$ MeV fm$^{-3}$ at high $L$. The variation with $p$ is more dramatic; for $p=4.0$ the difference with the baseline results is still relatively small ($<0.005$ fm$^{-3}$) but for $p=2.0$, $n_{\rm cc}$ and $P_{\rm cc}$ is significantly lower then the baseline results, falling by $\approx 0.02$ fm$^{-3}$. The effect of the variation of $p$ on $n_{\rm p}$ and $P_{\rm p}$ is still very small; the end result is that the region in which the pasta shapes are present is significantly reduced for $p=2.0$; the corresponding effect on macroscopic crust properties is discussed in section 7. \emph{The density region over which pasta is predicted to exist is thus very sensitive to the low proton fraction behavior of the surface energy.}

Finally, it can be seen that changing the magnitude of the surface tension for symmetric nuclear clusters $\sigma_0$ to 1.0 MeV fm$^{-3}$ has only a very small effect on the transition densities and pressures.

\subsubsection{Variation of SNM parameters}

Decreasing (increasing) $K_0$ and $n_0$ results in a decrease (increase) in $n_{\rm cc}$($P_{\rm cc}$) respectively. In Fig.~10 we plot the transition densities (b) and pressures (d) versus $L$ for $n_0=0.14, 0.17$ fm$^{-3}$ and $K_0=220,260$ MeV, compared to the baseline PNM sequence. Varying $n_0$ simply rescales the density dependence of the EoS; we thus plot the transition densities and pressures versus density scaled to the new saturation density ($\times n_0/0.16$fm$^{-3}$, which of course does not affect the results using $n_0=0.16$ fm$^{-3}$. Plotted in this way, $n_{\rm cc}$ and $P_{\rm cc}$ are elevated a small amount $\lesssim 0.005$ fm$^{-3}$ when $n_0$ is varied away from $0.16$ fm$^{-3}$. The lower (higher) $K_0$ values give lower (higher) values for $n_{\rm cc}$ and $n_{\rm p}$ by up to $\approx 0.005$ fm$^{-3}$; for the transition to pasta the effect is greatest at high $L$. Similarly, the pressures are lowered (raised) for lower (higher) $K_0$; the effect is greater at high $L$, up to $\approx 0.1$ MeV fm$^{-3}$ for $P_{\rm cc}$. Note that for the most part, $n_{\rm p}$ and $P_{\rm p}$ are quite insensitive changes in SNM properties.

\subsubsection{Variation of $K_{\rm sym}-L$ relations}

In Fig.~10 the results of our MSL baseline PNM sequence are compared to the equivalent sequence for the BD model. There is little difference between the MSL and BD results for the transition densities, but the PNM BD sequence gives significantly lower transition pressures than the equivalent MSL sequence, a result of the BD EoSs giving a consistently lower PNM pressure at intermediate sub-saturation densities than the MSL EoSs when constrained by the same low density PNM EoS. In addition, chemical equilibrium in the unit cell will require the nucleus or nuclear cluster to be larger and more diffuse to balance the reduced energy per particle of the free neutrons. This will reduce the pressure of the nucleus and hence the neutron gas still further. We can therefore expect that any quantities that are sensitive to the pressure and energies of the dripped neutrons will also be sensitive to differences in the sub-saturation form of the EoS.

\subsection{Pressure throughout the inner crust}

Fig.~11 shows the pressure scaled by $n^{4/3}$ over the inner crust for variations in surface energy parameters (left plot), SNM parameters (middle plot) and the low density form of the PNM EoS (right plot). We show results for the $L=25$ MeV (giving the lower pressure for $n \gtrsim 0.01$ fm$^{-3}$) and $L=70$ MeV (giving the higher pressure) members of the PNM sequences. From pressure equilibrium, the pressure will be dominated by the behavior of the energy per particle of the dripped neutrons - the PNM EoS.

\subsubsection{Variation of surface and SNM parameters}

As expected $P/n^{4/3}$ exhibits negligible dependence on the surface parameters (a), $n_0$ (once the density is rescaled by $n_0/0.16$fm$^{-3}$) and $K_0$ (b). Eqn~\ref{eq:eos4} shows a first-order dependence on $K_0$ but its effect is much smaller than that of $L$. 

\subsubsection{Variation of $K_{\rm sym}-L$ relations}

In Fig.~11c we show the $L=25$ MeV and $L=70$ MeV members of the BD PNM sequence compared with the members of the MSL PNM sequence with the same $L$. Because these members of the BD sequence have values of $J$ lower than the equivalent MSL members, we also show the $L=35$ MeV and $L=90$ MeV members of the sequence which have $J=25$ MeV and $J=35$ MeV respectively corresponding to the $J$ values of the members of the MSL PNM shown. The BD EoSs all differ significantly compared to the MSL equivalents in the higher density region of the crust, as the pressure from the BD EoSs is greatly reduced. We also plot the results from the widely used Douchin and Haensel \cite{Douchin2000} (DH) and Baym, Bethe, Pethick \cite{BBP1971} (BBP) EoSs, both of which have pressures above the MSL PNM sequence, but within the baseline shaded region of Fig.~7.

\begin{figure}[!t]\label{fig:11}
\begin{center}
\includegraphics[width=5.5cm,height=3.8cm]{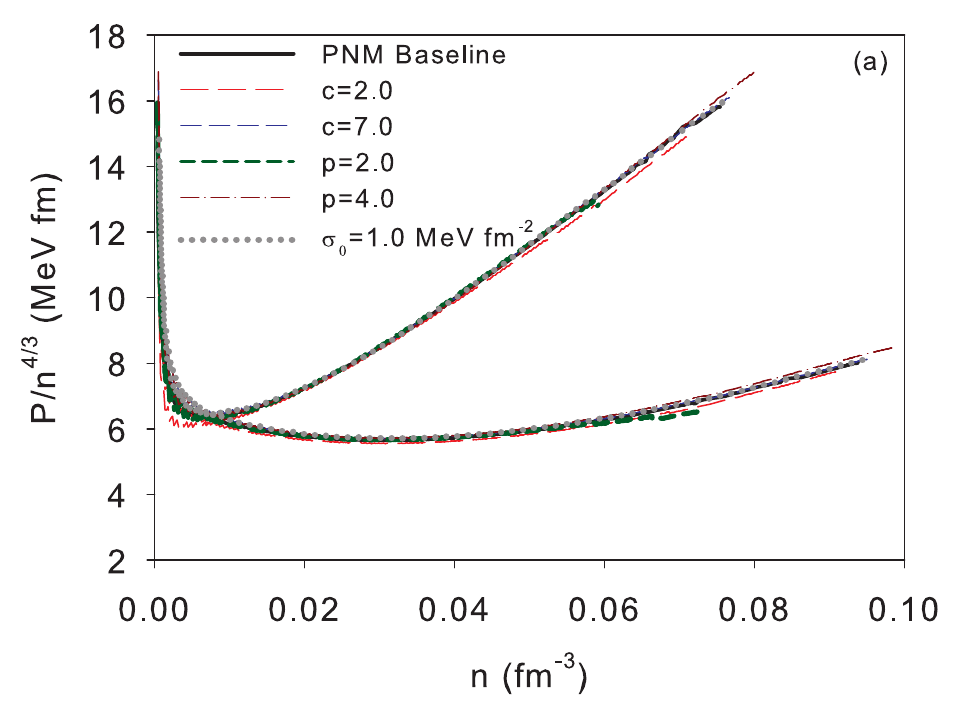}\includegraphics[width=5.5cm,height=3.8cm]{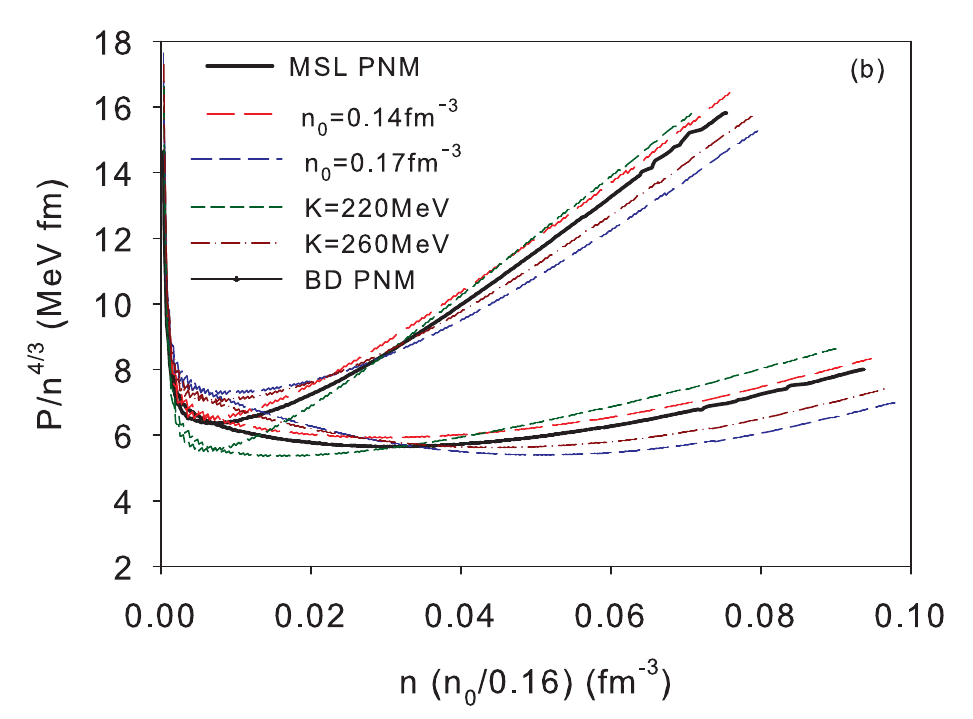}\includegraphics[width=5.5cm,height=3.8cm]{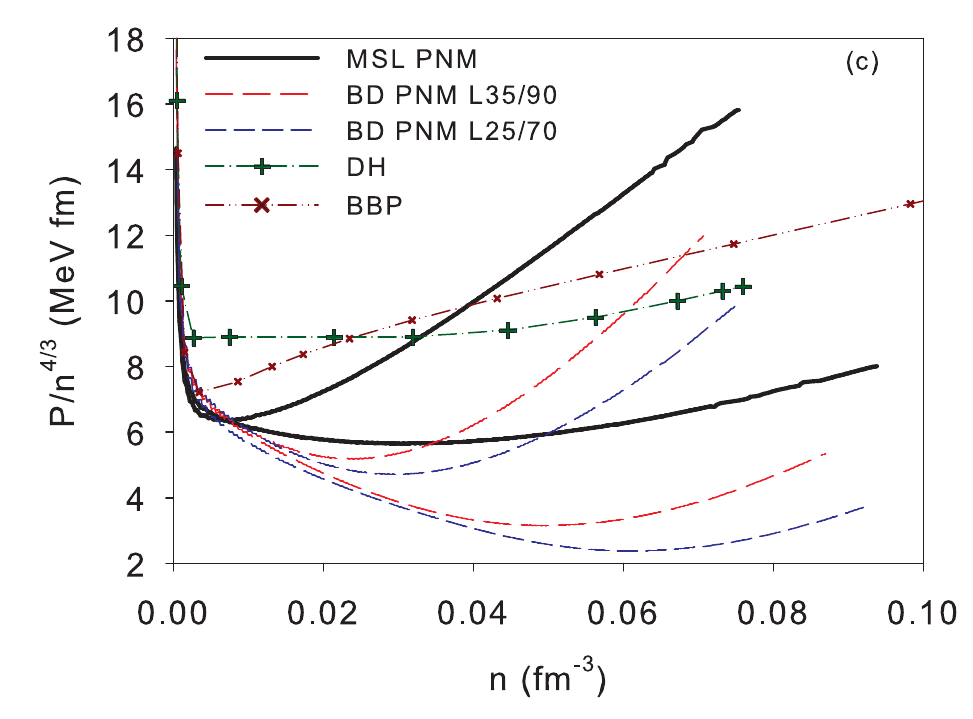}
\caption{(Color online) Pressure of matter scaled by $n^{4/3}$ versus baryon number density in the inner crust for different parameterizations of the surface energy (a), variations in the symmetric nuclear matter EoS (b) and for a different functional form of the nuclear matter EoS (BD) as well as two widely used liquid drop crust models (DH \cite{Douchin2001} and BBP \cite{BBP1971}) (c) compared to the baseline MSL PNM model. The higher (lower) values of $L$ correspond to the curves of higher (lower) pressure at $n \gtrsim 0.02$fm$^{-3}$.}
\end{center}
\end{figure}

\subsection{Volume and dripped neutron density fractions $v$, $X_{\rm n}$}

Fig.~12 shows the charged nuclear matter volume fraction $v$ (top plots) and dripped neutron density fraction $X_{\rm n}$ (bottom plots) over the inner crust for variations in surface energy parameters (a,d), SNM parameters (b,e) and the low density form of the PNM EoS (c,f). We show results for the $L=25$ MeV (giving the lower volume fraction and higher density fraction for $n \lesssim 0.06$fm$^{-3}$) and $L=70$ MeV members of the MSL PNM sequences. 

\begin{figure}[!t]\label{fig:12}
\begin{center}
\includegraphics[width=5.5cm,height=3.8cm]{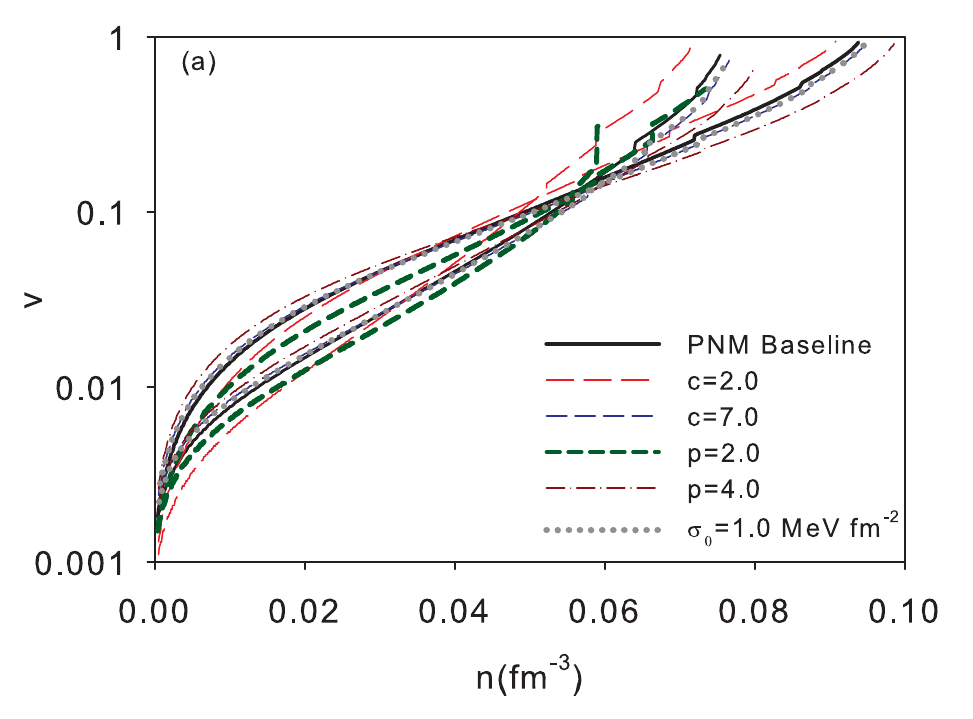}\includegraphics[width=5.5cm,height=3.8cm]{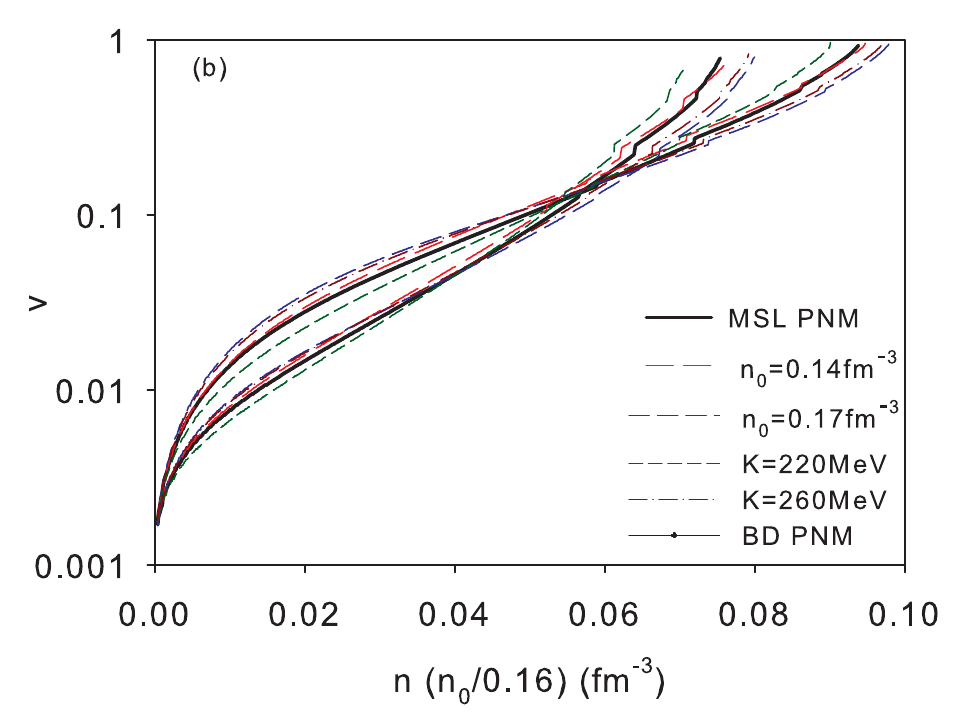}\includegraphics[width=5.5cm,height=3.8cm]{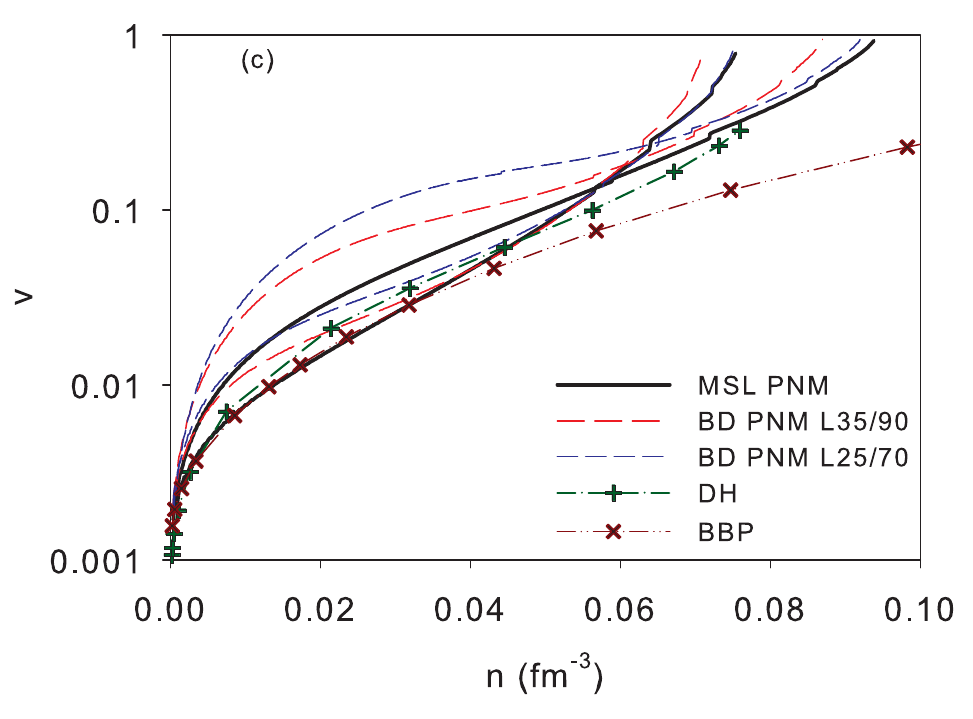}
\includegraphics[width=5.5cm,height=3.8cm]{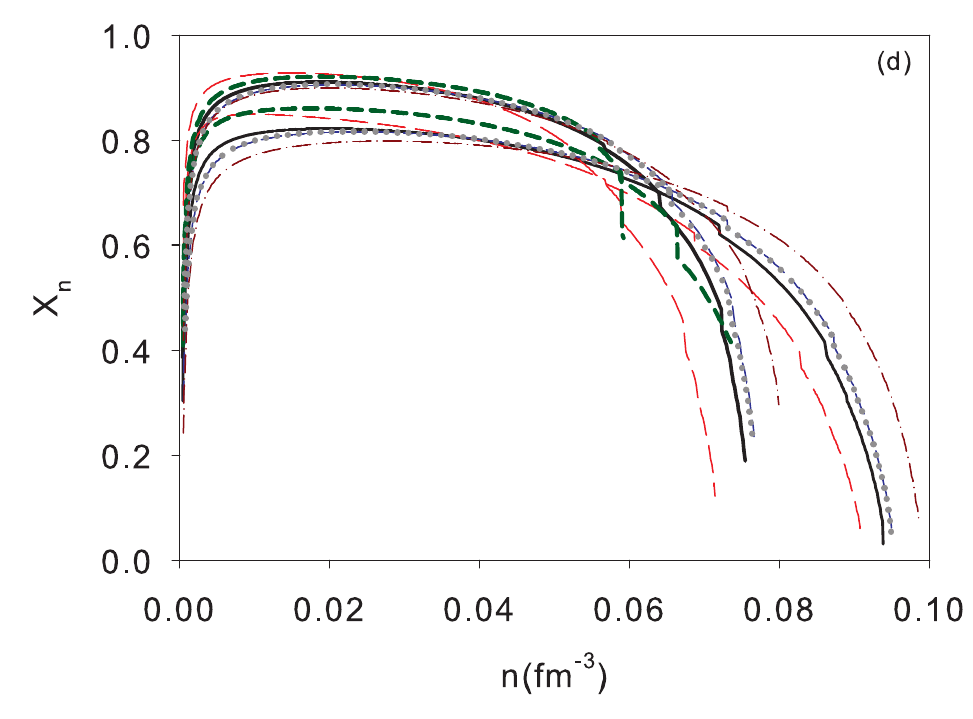}\includegraphics[width=5.5cm,height=3.8cm]{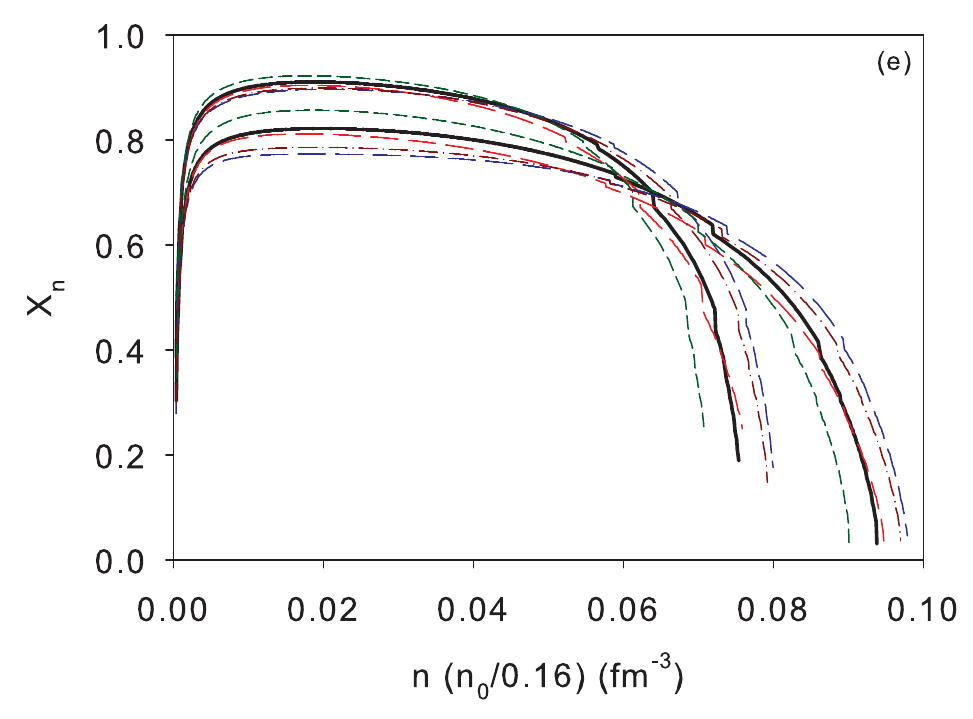}\includegraphics[width=5.5cm,height=3.8cm]{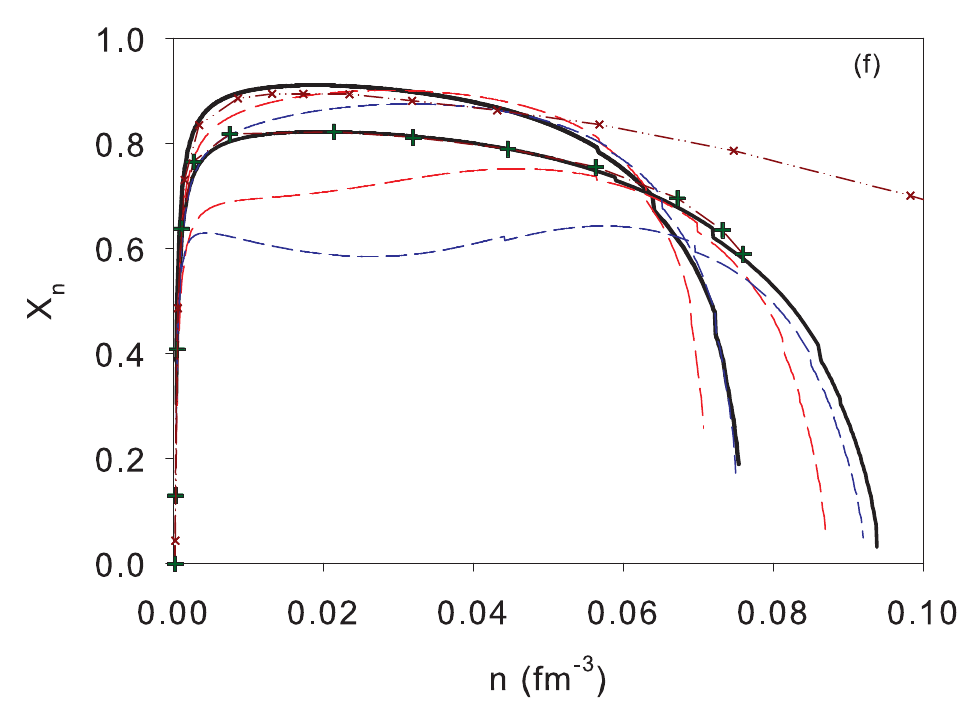}
\caption{(Color online) The volume fraction $v$ occupied by nuclei/nuclear clusters (a-c), and density fraction $X_{\rm n}$ occupied by free neutrons (d-f), versus baryon number density in the inner crust for different parameterizations of the surface energy (a,d), variations in the symmetric nuclear matter EoS (b,e) and for a different functional form of the nuclear matter EoS (BD) as well as two widely used liquid drop crust models (DH \cite{Douchin2001} and BBP \cite{BBP1971}) (c,f) compared to the baseline MSL PNM model. Results are shown for the $L=25$ and $70$ MeV members of the PNM sequences and in addition, for the BD EoS, the $L=35$ and $90$ MeV members. Lower $L$ gives the lower volume fraction and higher density fraction for $n \lesssim 0.06$fm$^{-3}$.}
\end{center}
\end{figure}

\subsubsection{Variation of surface and SNM parameters}

The volume fraction shows little dependence on $c$ throughout most of the crust; at higher densities it reaches the same maximum values as the baseline at a slightly lower (higher) density for $c$ = 2.0 (7.0). Throughout most of the crust, $p$ has little effect on the volume fraction; however, just as $p$ = 2 gives a large reduction in the transition density and pressure, the volume fraction at transition is also reduced significantly, down to about 0.3, reflecting the fact that the increased surface energy becomes prohibitive when the clusters occupy a much smaller fraction of the space. The free neutron fraction exhibits the same dependence on $c$ and $p$, with a significant effect only for $p$ = 2 at the crust-core transition density. $v$ and $X_{\rm n}$ show a very small dependence on $n_0$ and $K_0$ for the same reason;
that $v$ and $X_{\rm n}$ are largely determined by the PNM EoS.

\subsubsection{Variation of $K_{\rm sym}-L$ relations}

$v$ and $X_{\rm n}$ are very sensitive to the PNM EoS at intermediate sub-saturation densities, and hence to $K_{\rm sym}$ for fixed $L$.  For the reasons discussed in section~6.1.3, the BD EoS gives larger nuclei at a given density and thus a larger volume fraction of nuclei $v$ and smaller density fraction of dripped neutrons $X_{\rm n}$. This is particularly pronounced for lower values of $L=25,35$ MeV, for which the BD EoS predicts a reduction of $X_{\rm n}$ by about 0.2 and an increase in $v$ by a factor of $\approx 3$ for $n \lesssim 0.06$ fm$^{-3}$.

The DH EoS is also consistent, but the BBP EoS gives a much lower volume fraction and consequently much higher dripped neutron density fraction over the higher density region of the crust. This results from the high compressibility $K_0$ used for the BBP EoS and an unrealistically high surface tension.

\subsection{Local proton fraction $x$ and Wigner-Seitz cell radius $r_C$}

Fig.~13 shows the proton fraction in the nuclear clusters $x$ (top plots) and the Wigner-Seitz cell radius $r_C$ (bottom plots) over the inner crust for variations in surface energy parameters (a,d), SNM parameters (b,e) and the low density form of the PNM EoS (c,f). We show results for the $L=25$ MeV and $L=70$ MeV members of the MSL and BD PNM sequences and the $L=35$ MeV and $L=90$ MeV members of the BD PNM sequence in plots (c,f). The higher $L$ member gives higher proton fraction and lower cell radius.

\begin{figure}[!t]\label{fig:13}
\begin{center}
\includegraphics[width=5.5cm,height=3.8cm]{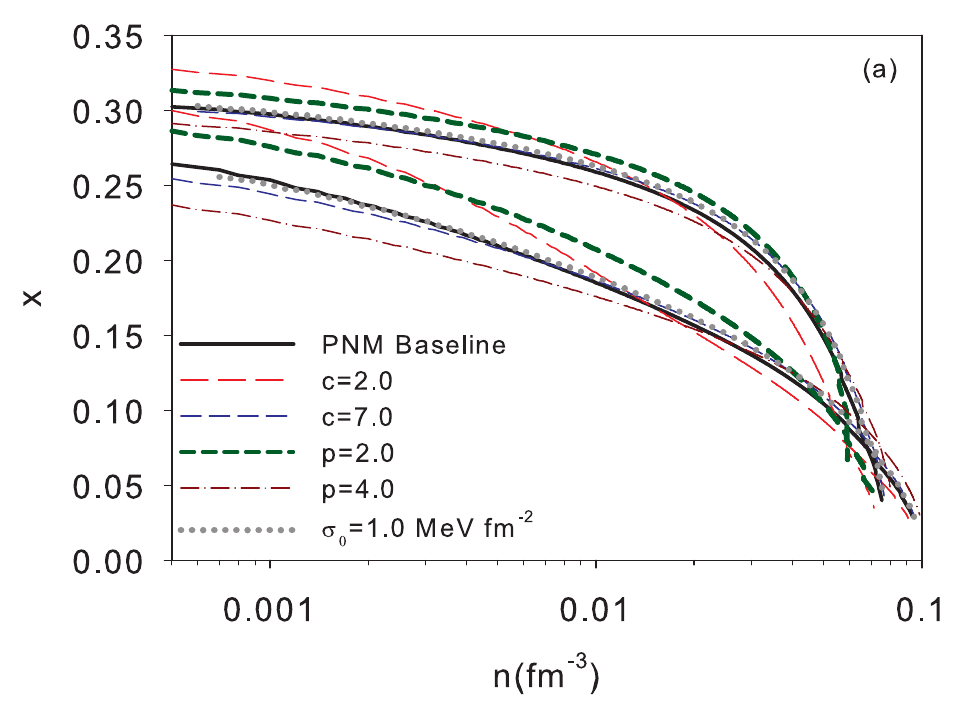}\includegraphics[width=5.5cm,height=3.8cm]{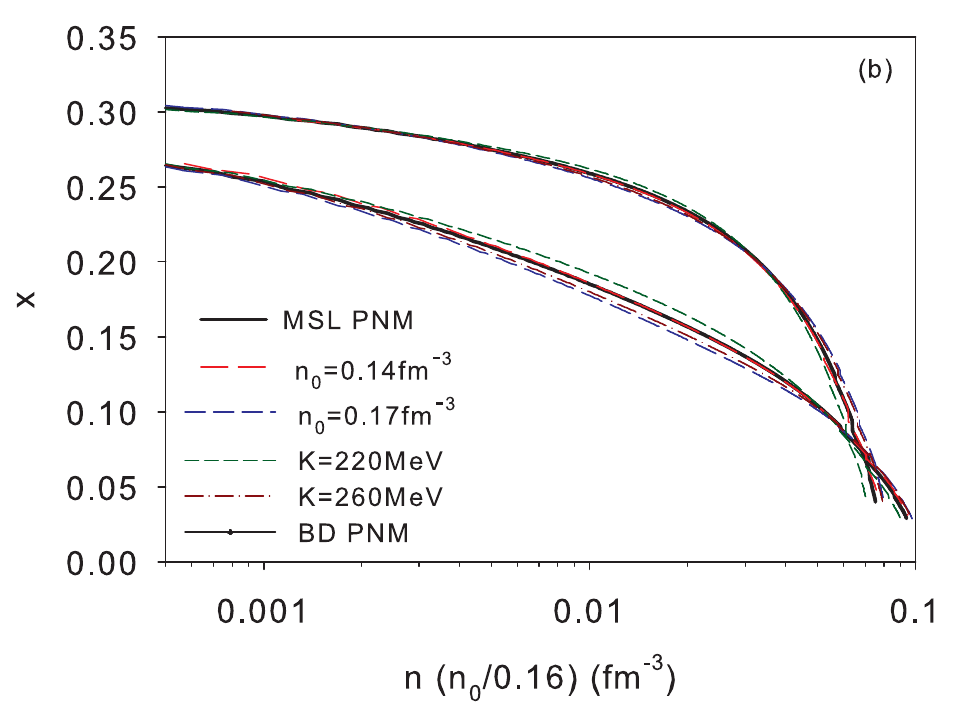}\includegraphics[width=5.5cm,height=3.8cm]{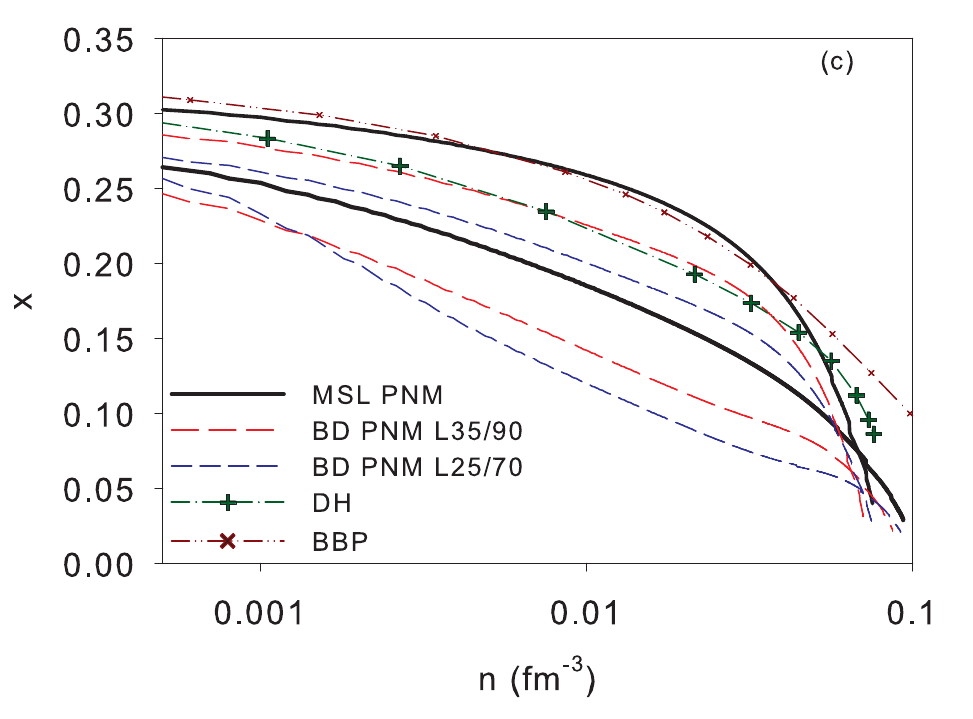}
\includegraphics[width=5.5cm,height=3.8cm]{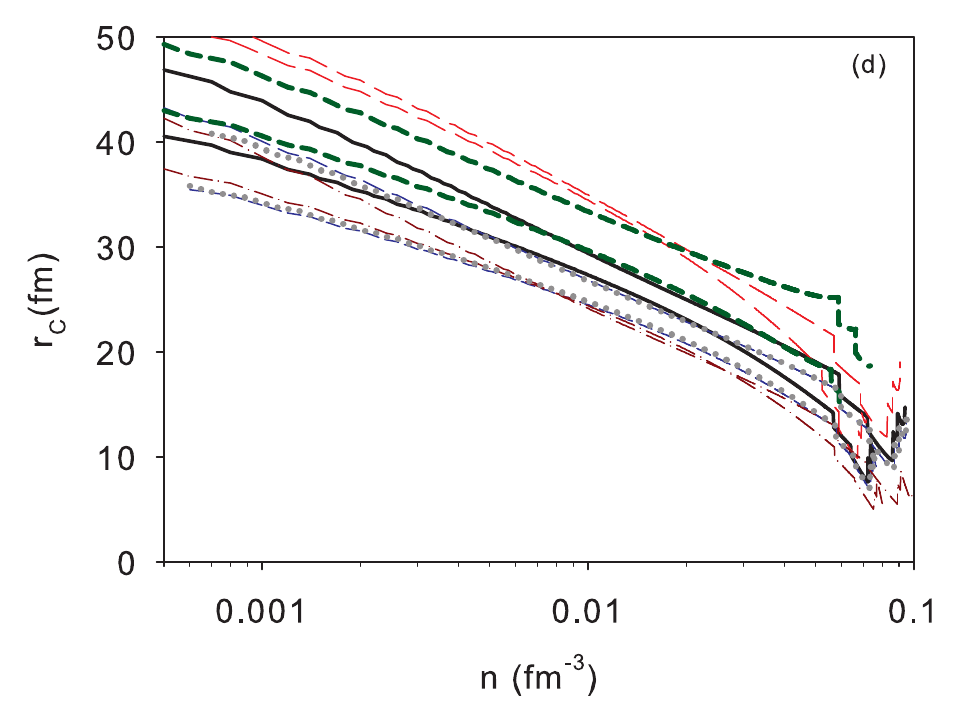}\includegraphics[width=5.5cm,height=3.8cm]{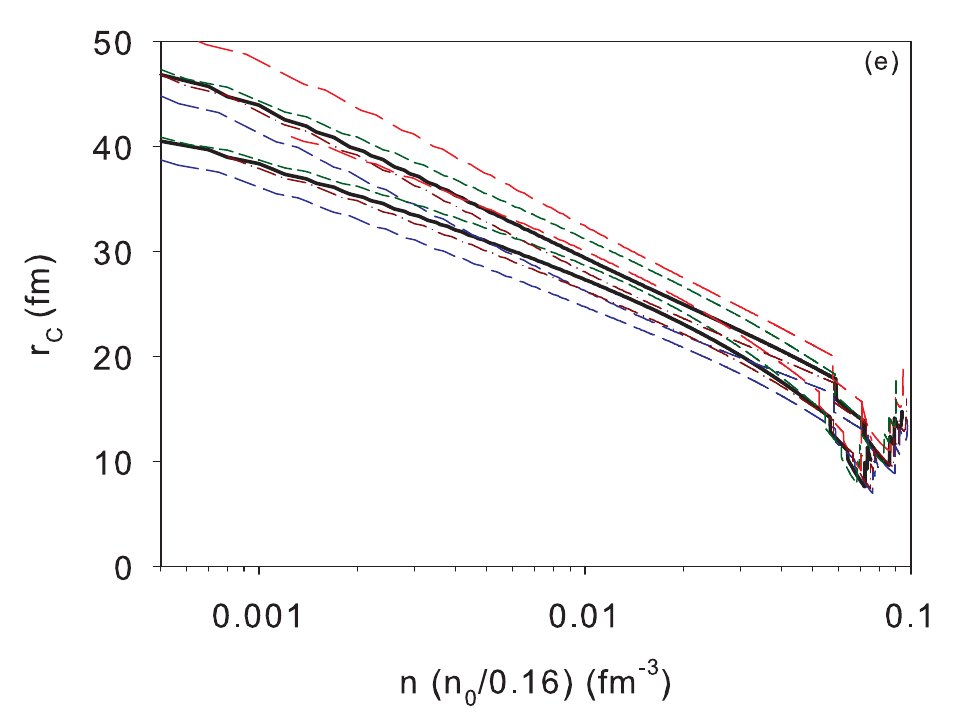}\includegraphics[width=5.5cm,height=3.8cm]{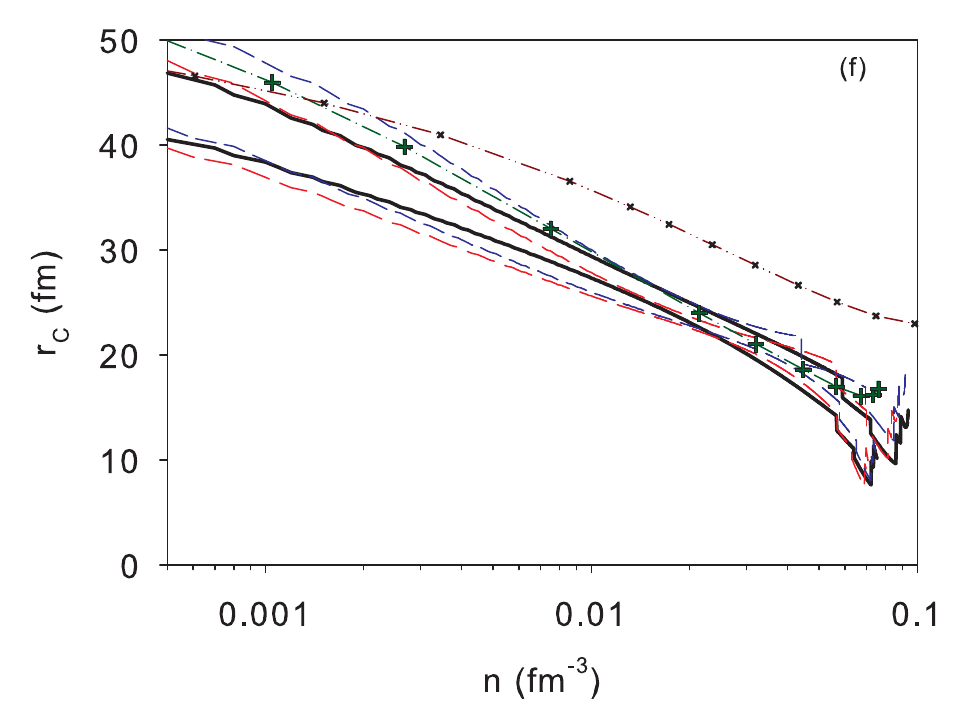}
\caption{(Color online) Local proton fraction of clusters $x$ (a-c) and WS cell size $r_{\rm C}$ (d-f) versus baryon number density in the inner crust for different parameterizations of the surface energy (a,d), variations in the symmetric nuclear matter EoS (b,e) and for a different functional form of the nuclear matter EoS (BD) as well as two widely used liquid drop crust models (DH \cite{Douchin2001} and BBP \cite{BBP1971}) (c,f) compared to the baseline MSL PNM model. The higher $L$ member of a sequence gives the curve with the higher proton fraction and lower cell radius.}
\end{center}
\end{figure}

\subsubsection{Variation of surface parameters}

$c$ and $p$ have relatively little effect on the proton fraction throughout most of the crust; a lower value of $c$ or $p$ (higher surface energy at high and low proton fractions) is associated with a slightly higher proton fraction at lower densities, as an increased proton fraction lowers the energy of the bulk matter inside the nuclei (makes it more bound) to compensate for the increased surface energy. In contrast, the size of the WS cell does exhibit a significant dependence on the surface parameters, shifting the baseline sizes on the order of 5 fm higher (lower) for lower (higher) values of $c$ and $p$ throughout most of the inner crust. Since the volume fractions are relatively unchanged, this means that the nuclear sizes are also higher (lower). This can be understood by considering the competition between the surface and bulk energies: raising the surface energy (by lowering $c$ or $p$) can be compensated by increasing the contribution to the total energy density of the bulk phases of matter relative to the surface contribution.

\subsubsection{Variation of SNM parameters}

$x$, determined in large part by the symmetry energy at a given density, shows negligible dependence on SNM parameters. The WS cell size $r_{\rm C}$ is relatively insensitive to $K_0$, but does show a significant dependence on $n_0$, increased by $\approx 5$ fm relative to the baseline results by lowering the saturation density to 0.14fm$^{-3}$.

\subsubsection{Variation of $K_{\rm sym}-L$ relations}

The WS cell size $r_{\rm C}$ appears insensitive to the reduced neutron pressure of the BD EoSs. In contrast, the local proton fraction $x$ drops by up to 0.05 for $L=25-35$ MeV; in order to reduce the lattice Coulomb energy cost of increasing the nuclear size (which happens because of the reduced dripped neutron pressure as outlined in section ~6.1.3), the proton fraction drops, which will maintain a roughly constant nuclear charge number $Z$. The DH EoS falls at the upper boundary of the WS cell size region. The BBP EoS predicts a larger WS cell size throughout most of the inner crust.

\section{Discussion}

\begin{figure}[!t]\label{fig:14}
\begin{center}
\includegraphics[width=8cm,height=6cm]{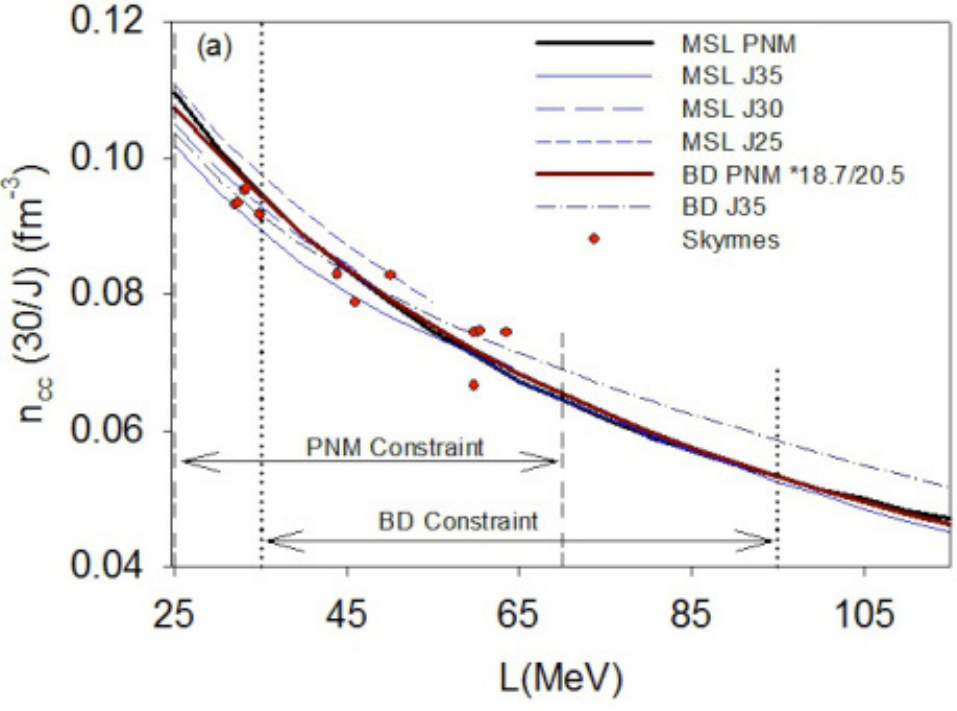}\includegraphics[width=8cm,height=6cm]{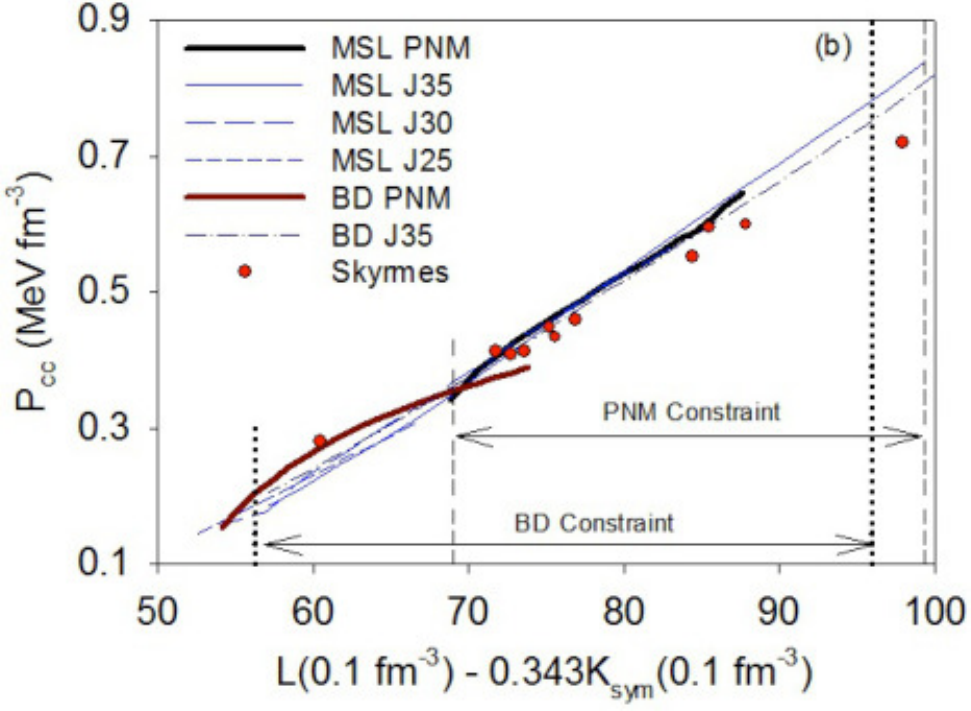}
\caption{(Color online) Crust-core densities versus $L$ scaled by 30 MeV/$J$ (a) and transition pressures versus $L(n=0.1$ fm$^{-3}$) - 0.343 $K_{\rm sym}(n=0.1 $fm$^{-3}$) (b) for the `PNM' and selected `J' sequences of MSL and BD EoSs, and for selected Skyrmes.}
\end{center}
\end{figure}

The transition densities for the sequence of constant $J$ EoSs (Fig.~6) indicates a roughly inverse linear scaling of the densities with $J$. With this in mind, we plot in the left panel of Fig.~14 the transition densities multiplied by $30/J$ (using $J=30$ MeV as a reference value). To add the `PNM' results we must of course use the correlation of $J$ with $L$.  We also add onto the plot the BD EoS results; for the BD PNM sequence we also scale the density according to the offset of the correlation relative to the MSL correlation. Once this is done, the results form a relatively tight relation with $L$. This relation is represented well by the quadratic fit

\be
n_{\rm cc} (30/J) = 0.135 - 0.098 (L/70) + 0.026 (L/70)^2 \pm 0.005
\ee 

\noindent where the dominant error in the fit comes from the $1\sigma$ bounds on the constant term.

It was suggested that a tighter correlation of the transition pressure with density would be obtained by plotting $P_{\rm cc}$ against some combination of EoS parameters rather than just $L$ \cite{Ducoin2011}; an optimum fit was obtained for $L(n=0.1 $fm$^{-3}$) - 0.343 $K_{\rm sym}(n=0.1 $fm$^{-3}$) using transition pressures obtained using dynamical and thermodynamical stability analyses. The pressure of the matter at the crust-core transition is related to the pressure of PNM at around the same density ($\sim 0.1$ fm$^{-3}$); from Eqn.~\ref{eq:eos5}, this will be related, to first two orders, with $L$ and $K_{\rm sym}$ at  $n \sim 0.1$ fm$^{-3}$ when $K_0$ is fixed, so the correlation obtained is not surprising. We plot the same quantities obtained in our model for the PNM and constant $J$ sequences of both MSL and BD EoSs, as well as the Skyrme EoSs, on the right panel of Fig.~14. A tight correlation is indeed observed. This emphasizes the fact that more accurate constraints on both $L$ \emph{and} $K_{\rm sym}$ at sub-saturation densities are required to improve our estimate of the NS crust transition pressure. Our relation is well fit by

\be
P_{\rm cc} = -0.724 + 0.0157[L(0.1 \text{fm}^{-3}) - 0.343 K_{\rm sym}(0.1\text{fm}^{-3})],
\ee

\noindent compared with the relation obtained in \cite{Ducoin2011} of $P_{\rm cc} = -0.328 + 0.0959[L(0.1$fm$^{-3}$) - 0.343 $K_{\rm sym}(0.1$fm$^{-3})]$. The differences likely arise from the method used to locate the crust core transition density; in  \cite{Ducoin2011} it is taken to be the density location of the edge of the thermodynamic spinodal of beta-equilibrated nuclear matter, below which matter becomes unstable to clustering.

The regions over which the MSL and BD EoSs are consistent with both the low density PNM constraint and $25<J<35 $ MeV are depicted on the plots; these are still relatively large allowed regions. For MSL, the region spans transition densities $\approx 0.076 - 0.12$ fm$^{-3}$ and transition pressures  $\approx 0.36 - 0.8$ MeV fm$^{-3}$; for BD the regions are $\approx 0.07 - 0.11$ fm$^{-3}$ and $\approx 0.2 - 0.75$ MeV fm$^{-3}$ respectively. The differences in the two ranges reflects the uncertainty in $K_{\rm sym}$ and hence the uncertainty in the sub-saturation behavior of the PNM EoS. The Skyrmes all fall within the MSL range as expected, since the MSL model mimics the sub-saturation behavior of the Skyrme EoSs.

\begin{figure}[!t]\label{fig:15}
\begin{center}
\includegraphics[width=5.5cm,height=4.3cm]{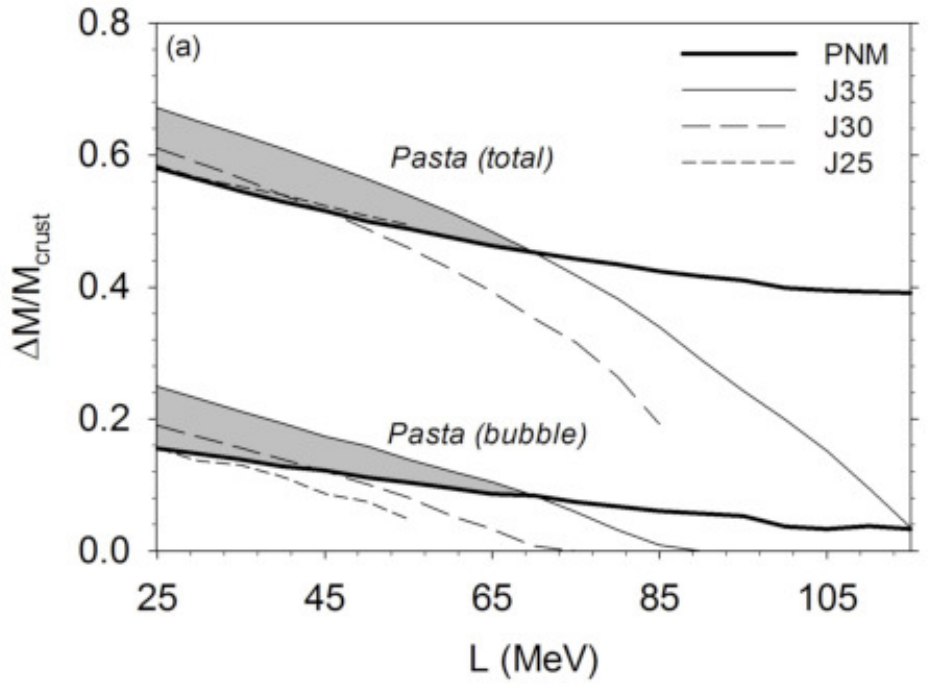}\includegraphics[width=5.5cm,height=4.3cm]{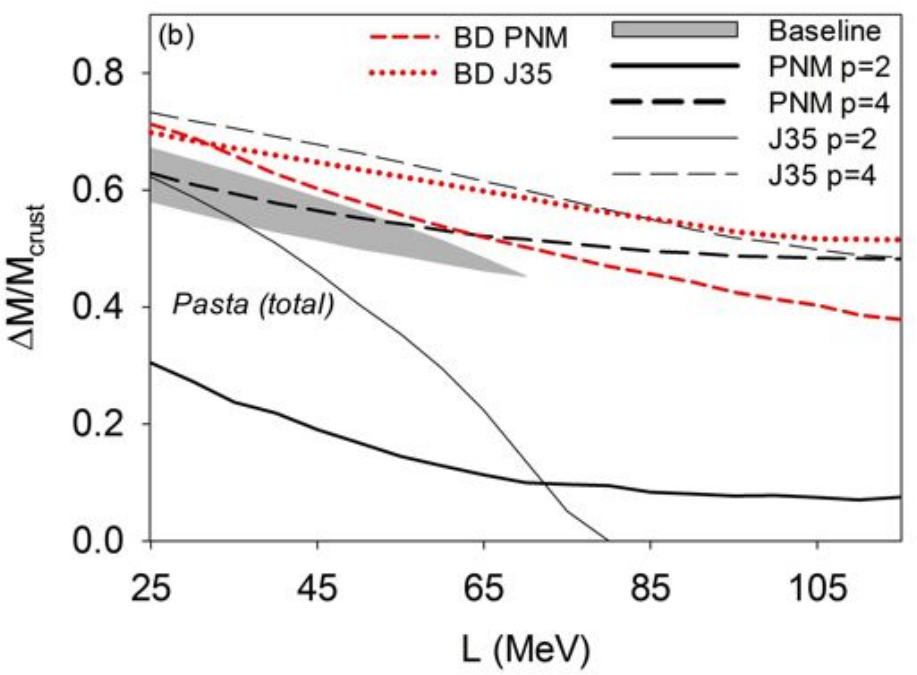}\includegraphics[width=5.5cm,height=4.3cm]{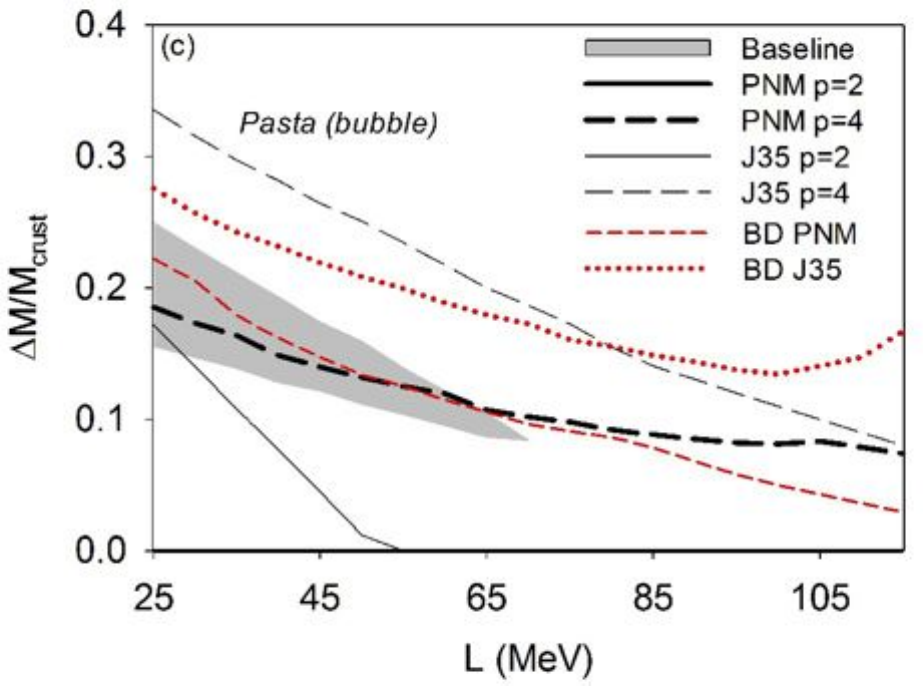}
\caption{(Color online) Fraction of mass contained within the pasta phases as a whole and the bubble phases relative to the total crust mass versus $L$. We show the baseline results (a) and the variation with the surface parameter $p$ (c,d).}
\end{center}
\end{figure}

The transition pressures can be used to estimate the mass and moment of inertia fractions of various components of the crust relative to the crust totals. Assuming the outer layer of the neutron star to be thin and contain relatively little mass compared to the whole star, the mass of that layer is proportional to the pressure at the base of that layer $\Delta M \propto P_{\rm base}$ \cite{LorenzPethick1993}, otherwise depending only on the bulk properties of the star. From this, the fraction of mass contained within the pasta phases, and within only the bubble pasta phases, can be estimated as
\be
{\Delta M_{\rm pasta} \over \Delta M_{\rm crust}} \approx 1 - {P_{\rm p} \over {P_{\rm cc}}}; \;\;\;\;\;\;\;\;\; {\Delta M_{\rm bubble} \over \Delta M_{\rm crust}} \approx 1 - {P_{\rm bu} \over {P_{\rm cc}}}.
\ee
\noindent where $P_{\rm bu}$ is the pressure at the transition to the bubble phases of pasta. Since the moment of inertia contained within the various phases relative to the total crustal moment of inertia is proportional to $\Delta M$, the moment of inertia fractions will be identical to the mass fractions.

We plot the mass fractions in Fig.~15. In the left plot, the baseline MSL results are shown. For $L<70$ MeV, the pasta phases account for a large fraction of the mass of the crust: between 50\% and 70\%. Although the pasta layers occupy a relatively thin layer in terms of thickness, the density of those layers is the highest in the crust, resulting in potentially large contributions to the overall mass of the crust. For $L>70$ MeV, the mass fraction levels out for the PNM sequence, always above 40\%, while for the J35 sequence it falls off rapidly to just a few percent at $L=$ 115 MeV.

Even the bubble phases of pasta can occupy a substantial portion of the crust; between 10\% and 25\% for $L<70$ MeV, remaining above 10 \% for the PNM sequence and dropping to zero at $L\approx85$ MeV for the J35 sequence. In the bubble phases the protons are no longer localized in space, opening up a range of possible effects such as the direct Urca process \cite{Gusakov2004} and bulk super-flow of protons. The possibility of a significant extent to the bubble phases thus has implications for cooling and the dynamics of the crust-core interface.

As we have seen, the transition pressures are most affected by variations in the behavior of the surface energy at low proton fraction, as explored through the variation of the parameter $p$, and through variation in the sub-saturation behavior of the PNM EoS while keeping saturation properties constant. In the middle and right panels, the baseline results are compared with the results using $p=2$ (stiffer surface energy at low proton fraction) and $p=4$ (softer), as well as with the BD EoS. For $p$ = 4, the mass fractions of pasta and the bubble components are elevated slightly compared to the baseline. For $p$ = 2 the mass fractions are greatly reduced; remaining between 10\% and 30\% for the total pasta fraction on the PNM sequence, and falling from 60\% at $L=$ 25 MeV down to 0\% at $L=$80 MeV on the `J35' sequence. The bubble phases almost completely disappear for $p=2$, existing only for $L<50$ MeV for any sequence; the bubble fractions are raised for $p=4$, accounting for 10\%-35\% of the crustal mass.

The mass fractions of the total pasta and bubble phases for the BD EoS are elevated relative to the MSL results; the total pasta fraction remaining above 60\% for the BD J35 sequence, and above 40\% for the PNM sequence, while the BD J35 sequence gives rise to a crust in which more than 20\% of the mass is contained within the bubble phases. It is important to note that this is consistent with the fact that the transition pressures are much lower for the BD EoSs; the transition pressures set the absolute mass of the crust, which is lower for the BD models; however, the fraction of that mass in pasta and in bubbles is higher. 

We now give some simple examples of how some bulk crust quantities of interest in physical models of neutron star phenomena are affected by the previous exploration of the CLDM. We will focus on the charge and mass number of nuclei, the shear modulus of the crust and its melting temperature, displayed in Fig.~16. All these quantities are defined only within the region of the crust in which separate nuclei exist and become undefined or the expression we use is invalid in the pasta phases.

\begin{figure}[!t]\label{fig:16}
\begin{center}
\includegraphics[width=7.8cm,height=5cm]{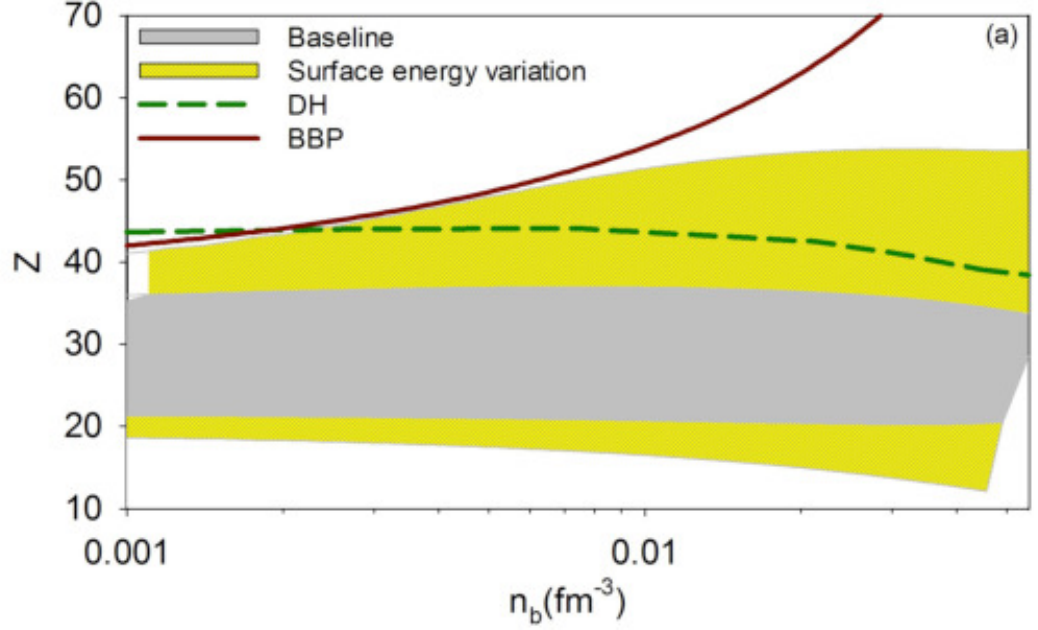}\includegraphics[width=8cm,height=5cm]{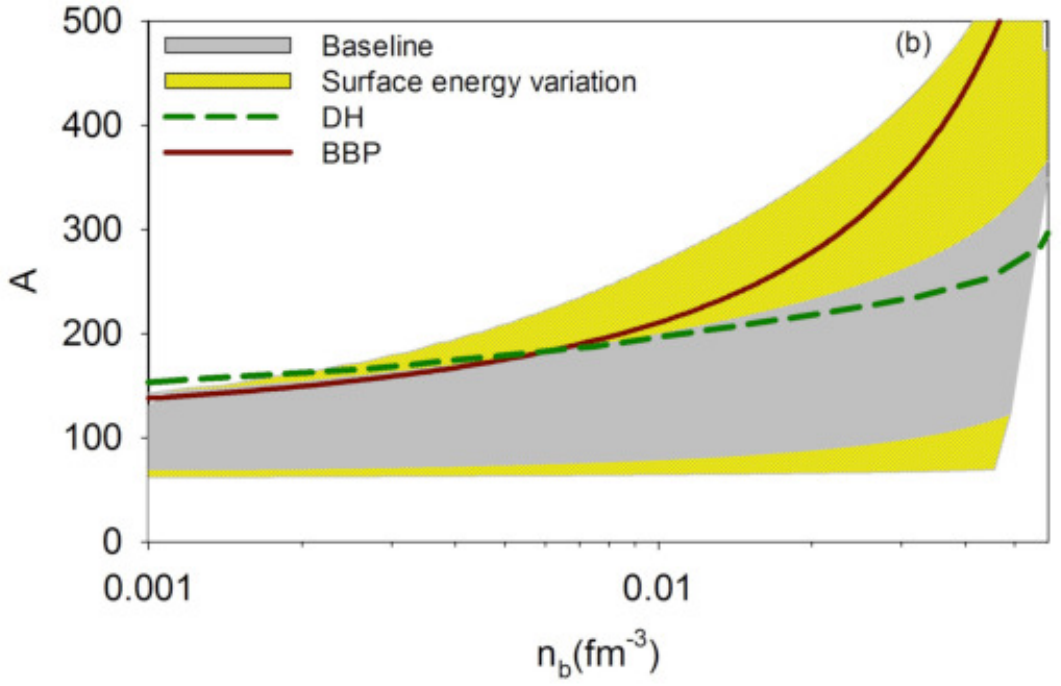}
\includegraphics[width=8cm,height=5cm]{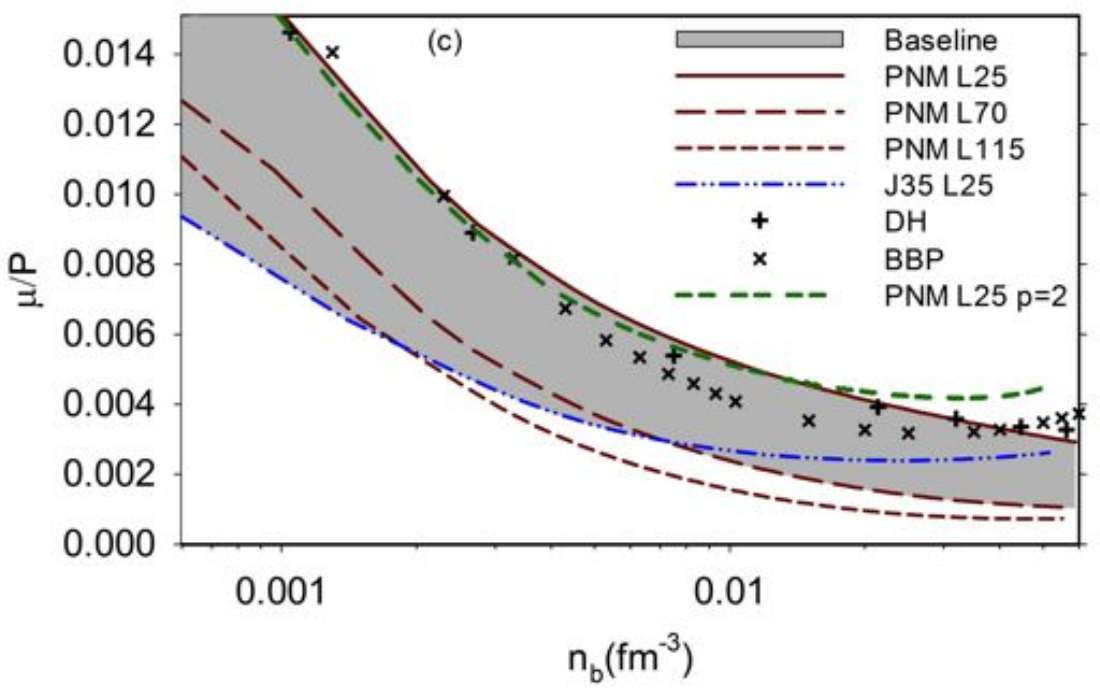}\includegraphics[width=8cm,height=5cm]{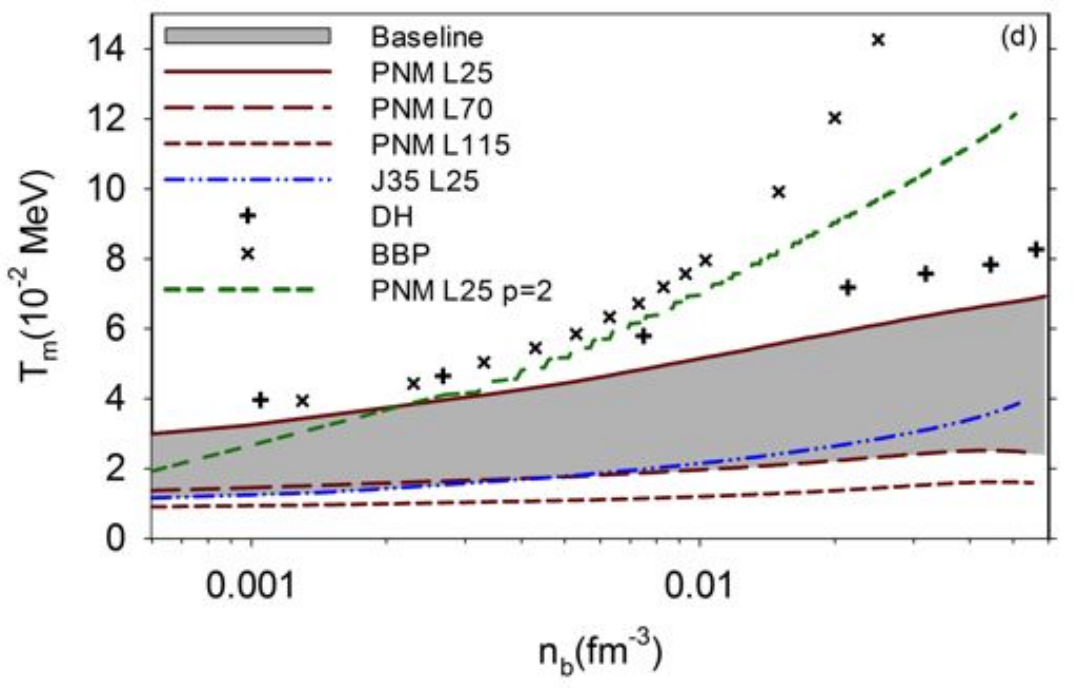}
\caption{(Color online) Top: Charge $Z$ (a) and mass $A$ (b) number of nuclei in the crust up to the transition to pasta versus density for the baseline MSL EoSs (shaded area) and the variation with surface energy (hatched area) together with the results from the Douchin and Haensel EoS \cite{Douchin2001} (DH) and BBP EoS \cite{BBP1971} (BBP). Bottom: shear modulus of crust scaled by pressure (c) and melting temperature of crust (d) versus density for the baseline MSL EoSs.}
\end{center}
\end{figure}

The top left and right panels show $Z$ and $A$ over the inner crust. The baseline results are the shaded band; for a given EoS $Z$ tends to remain roughly constant, and varies by a factor of 2 from 20 to 40. Increasing the surface tension at high and low proton fractions tends to increase the nuclear size to reduce the surface energy compared to the bulk; thus $Z$ increases. The hatched band takes into account those variations through $c$ and $p$ and increases $Z$ up to about 50 in the higher density regions of the crust. The DH EoS lies along the upper bound of the baseline results; BBP diverges to much higher $Z$ at higher densities, an artifact of the much higher surface energy in that model. The mass number $A$ shows a similar trend: the baseline results predict $A$ to rise from between 60 and 150 at the lowest densities; higher $L$ predicts a slower rise. Increasing the surface energy increases $A$ and the rate at which it increases, reaching towards 1000 at the transition to pasta. Again, DH runs along the upper bound of the baseline results, while BBP predicts much higher $A$ at high densities, in line with our results for high surface energy.

The shear modulus of a Coulomb lattice of positively charged nuclei in a uniform negatively charged background in the NS crust at a baryon number density $n_{\rm b}$ was determined through Monte-Carlo simulation \citep{Ogata1990, Strohmayer1991, Chugunov2010} and can be written as
\be \label{eq:shear_mod}
	\mu = 0.1106 \left(\frac{4\pi}{3}\right)^{1/3} A^{-4/3} n_{\rm b}^{4/3} (1-X_{\rm n})^{4/3} (Ze)^2,
\ee
\noindent where the nuclei are characterized by the nucleon and proton number $A,Z$ and $X_{\rm n}$ is the fraction of neutrons not confined to the nuclei. This is the zero-temperature expression, a good approximation for neutron star temperatures below $\sim 10^8$K. Technically this is valid only when the ions (nuclei) can be treated as point charges in a uniform background of electrons; finite size effects and the dripped neutrons are not taken into account. Nevertheless, it is widely used to approximate the shear modulus of the inner crust.

The melting temperature (at which the crystalline lattice changes to a gas of ions) can be expressed as \cite{HaenselBook2007}
\be \label{eq:melting_temp}
T_{\rm m} = { (Ze)^2 \over 175 k_{\rm B} r_{\rm C}}
\ee
\noindent neglecting quantum effects like zero-point vibrations.

The bottom two panels of Fig.~16 show the shear modulus scaled to pressure (left) and the melting temperature (right) relative to 0.01 MeV ($1.16 \times 10^8$ K). The baseline results for the shear modulus show a variation of a factor of about 2 throughout the inner crust; the melting temperature varies by a factor of 3 throughout the crust. The shear modulus appears relatively insensitive to variations in the surface energy: we show by way of comparison the MSL EoS with the surface parameter $p$ = 2 which gives the largest deviations in the transition densities and composition compared to the baseline results.

%
%

\section{Conclusions}

Using a modified Skyrme-like (MSL) nuclear matter model, we have constructed a sequence of baseline inner crust EoSs using the compressible liquid drop model (CLDM), spanning a conservative estimation of the experimentally constrained range of the slope of the symmetry energy at saturation density $25<L<115$ MeV. The magnitude of the symmetry energy at saturation density, $J$, was then determined two ways: from the requirement that the low density PNM EoS be consistent with that derived from recent, robust, ab-initio calculations (leading to what we refer to as the PNM sequence of crust models) and by holding J constant at a specific value in the range 25 - 35 MeV (what we refer to as the J25 - J35 sequences of crust models). 

Using these sequences as baseline models for the crust, we then examined the range of predictions of inner crust properties around the baseline models by varying the symmetric nuclear matter parameters and, in a model independent way, the surface energy parameters, within reasonable bounds informed by experiment and general results from theoretical modeling of the nuclear surface. By also constructing a similar set of inner crust models using a different functional form of the nuclear matter EoS (the Bludman-Dover (BD) EoS), we also explore the sensitivity of crustal quantities to $K_{\rm sym}$ which gives rise to differences in the pure neutron matter (PNM) EoS at sub-saturation densities for fixed saturation density parameters.

Consistency of the PNM constraint with the range $J$ = 25-35 MeV constrains $L<$70 MeV in the MSL functional; the sequences of crust models consistent with these conditions give the following ranges for crustal parameters:

\begin{itemize}\addtolength{\itemsep}{-0.5\baselineskip}
\item{$n_{\rm cc}$ =  0.08-0.12 fm$^{-3}$; $n_{\rm p}$ = 0.05-0.06 fm$^{-3}$. Pasta phases exist to some extent throughout the whole range $25 < L < 115$ MeV. The correlation of $n_{\rm cc}$ and $n_{\rm p}$ with $L$ is similar to those obtained by previous studies \cite{Oyamatsu2007,Xu2009}, but additionally we have shown that the slope of the correlation of  $n_{\rm cc}$ with $L$ is quite sensitive to the correlation of the $L$ with $J$ in a particular model. By scaling $n_{\rm cc}$ to the symmetry energy $J$, the correlations with $L$ fall closely along the same curve $n_{\rm cc} (30/J) = 0.135 - 0.098 (L/70) + 0.026 (L/70)^2 \pm 0.005$, independent of the EoS used.}

\item{$P_{\rm cc}$ =  0.35-0.8 MeV fm$^{-3}$; $P_{\rm p}$ = 0.15-0.3 MeV fm$^{-3}$. We have extended the results of  \cite{Ducoin2011} to show within the CLDM that $P_{\rm cc} = -0.724 + 0.0157[L(0.1 \text{fm}^{-3}) - 0.343 K_{\rm sym}(0.1\text{fm}^{-3})]$ independent of $J$ and EoS used.}

\item{The pressure throughout the crust varies by $5 < P/n^{4/3} <15$ MeV fm.}

\item{The volume fraction of the charged component of the crust $v \approx$ 0.001 - 0.15 in the region $n$ =  0.001fm$^{-3}$ up to the transition to the pasta phases $n_{\rm p}$, rising to the crust-core transition volume fraction of between 0.8 and 0.96.}

\item{The density fraction of neutrons in the pure neutron fluid, $X_{\rm n}$, rises rapidly from 0 at the neutron drip point to between 0.7-0.9 in the range  $n$=0.005 fm$^{-3}$-$n_{\rm p}$ ; it then falls to its transition value of between 0.01 and 0.2.}

\item{The Wigner-Seitz cell radius/half-width $r_{\rm C}$ can be roughly be described by the relation $r_{\rm C} \approx -14.5 \log{n} - 3.5 \pm 4$ fm throughout the crust.}

\item{The proton fraction in the charged nuclear component $x$ can roughly be described by the relation $x \approx -2n + 0.24 \pm 0.03$ in the range $n=0.01 - 0.1$ fm$^{-3}$; at lower densities $x$ rises to $0.3 \pm 0.05$ at neutron drip density.}

\item{The fraction of the total crust mass containing the pasta phases lies in the range 50 - 70\%. The mass fractions of the bubble phases, in which the protons are free, is in the range 10 - 25\%.}

\item{The charge and mass numbers $Z$ and $A$ fall in the ranges $Z$ = 20-35 throughout most of the crust, with the lower bound dropping to $Z$=10 at the transition to pasta, and $A$ = 60-200 throughout most of the crust, with the higher bound rising to 300 at the transition to pasta.}
\end{itemize}

The above ranges are greater when one relaxes the requirement of $L<70$ MeV. Note that whilst the above ranges give a good guide to the range of crustal parameters, the dependence on the symmetry energy within those ranges is more complicated, and to probe the effect of the symmetry energy and PNM EoS on neutron star phenomena, \emph{one should use crustal EoSs consistent with the core EoSs based on the symmetry energy parameters}.

Of the other model parameters, the inner crustal properties were found to be most sensitive to the low proton fraction behavior of the surface energy (controlled by the parameter $p$) and the behavior the PNM EoS at densities $\sim 0.02$ fm$^{-3}$ - $n_{\rm 0}$ for fixed $L$ and low density PNM EoS.

We show that decreasing $p$ from 3.0 (the baseline) to 2.0 significantly reduces the baseline $n_{\rm cc}$ range to 0.06 - 0.11 fm$^{-3}$ and $P_{\rm cc}$ to 0.2 - 0.75 MeV fm$^{-3}$. The mass fraction of pasta in the crust is reduced to 0.1 - 0.6 for $L<70$ MeV, vanishing completely for $L>70$ MeV in the models with constant $J$. The effect of the surface symmetry-bulk symmetry slope parameter $c$ is much less pronounced; decreasing $c$ from the baseline value to the minimum value used, 2, reduces $n_{\rm cc}$ by $\approx 0.005$ fm$^{-3}$ and $P_{\rm cc}$ by $\approx$ 0.05 MeV fm$^{-3}$. Increasing (decreasing) the surface tension was found to alter $Z$ by up to +10 (-5) throughout the crust. 

The composition is relatively unaffected by changes to $c$ and $p$ except for the WS cell size $r_{\rm C}$ which increases by $\approx5$ fm by decreasing either $c$ or $p$ to their minimum values (that is, increasing the strength of the surface and curvature tensions to their maxima). The saturation density $n_0$ has little effect on any of the crustal properties when the densities are scaled to $n_0$, except for $r_{\rm C}$ which increases by $\approx5$ fm when lowering $n_0$ to 0.14 fm$^{-3}$. Changing $K_0$ in the range 220-260 MeV has a small effect on the transition properties, raising or lowering $n_{\rm cc}$ by $\pm 0.003$ fm$^{-3}$, with negligible effect on the crustal composition.

The PNM sequence of models from the BD EoS predict higher values of the symmetry energy curvature at saturation density $K_{\rm sym}$ and hence lower energy and pressure for PNM than the MSL sequences (Fig.~1b). This leads to a significantly lower crust-core and spherical nuclei-pasta transition pressure and a higher volume fraction of nuclei, lower density fraction of dripped neutrons and lower local proton fraction throughout the crust. One should note, however, that the BD EoSs cannot match the latest ab-initio calculations of PNM within estimates of uncertainty in the three-body interaction (GCR and HS in Fig.~1) through the whole sub-saturation density regime; the MSL EoSs pass through those bounds for a limited range of $40 \lesssim L \lesssim 70$ MeV, a range also consistent with Skyrme and RMF models constrained by PNM calculations \cite{Fattoyev2012}. However, sensitivity to higher order symmetry energy parameters illustrates the need to constrain $K_{\rm sym}$ to within tighter limits in addition to $L$ in order to better estimate the transition densities and composition of the crust.

Thus, one main conclusion is that in order to constrain the crust composition and transition properties further, the most urgent work must be done in constraining the PNM EoS at sub-saturation densities (equivalently, $K_{\rm sym}$) and the surface tension of nuclear clusters with low proton fractions. Ab-initio PNM calculations are now sufficiently robust, particularly at low densities, that they offer constraints on the EoS as rigorous as those coming from nuclear experiment; we contend that the PNM sequences of inner crust models are a more compelling set of predictions than those sequences in which $J$ is held constant. Recent calculations of neutron drops suggest that the the surface energy of neutron drops, and thus low proton fraction clustered matter, might be under-predicted from fits to calculations involving Skyrme functionals \cite{Gandolfi2_2011}.  In our model, this corresponds to lower $p$, and as we have shown, this would predict a significantly smaller, possibly vanishing layer of pasta in the crust, potentially altering its properties appreciably.

Our results are consistent with other widely used crustal EoSs such as DH. Comparing their EoS with the results we obtain using the SLy4 Skyrme (as used in DH), with the baseline surface parameters, shows that DH obtain a larger $r_{\rm C}$ (still within our overall baseline region) and $Z$. Additionally, DH predicts no pasta, whereas we obtain a sizable pasta region for the same Skyrme parameterization and the baseline surface parameters. This suggests that DH obtained a stiffer surface energy at low proton fractions than our baseline surface energy, and emphasizes the sensitivity of the CLDM to this high isospin asymmetry surface behavior. Without knowing the exact parameters used in that model, however, we are confined to speculation.

We have shown that properties of the crust important in calculating the hydrodynamics and thermodynamics of neutron stars (e.g. shear modulus), as well as the thickness, mass and moment of inertia fractions of the crust, are sensitive to the EoS of uniform, neutron rich matter. The uncertainties in the saturation and sub-saturation behavior of the symmetry energy (or equivalently the PNM EoS) and the nuclear surface energy at high isospin asymmetry lead to the greatest uncertainties in transition densities and crust composition. Coupled with the sensitivity of global neutron star properties to the same nuclear EoS properties, models of phenomena with potentially observable signatures such as neutron star oscillations, heating and cooling, crustal deformations and the dynamical couplings between crust and core can be quite sensitive to nuclear surface properties, the symmetry energy and PNM EoS. Such sensitivities may only appear significant when crust and core descriptions are consistent \cite{Gearheart2011}. To that end, the sequences of crust EoSs discussed in this paper are available for use (\url{http://williamnewton.wordpress.com/ns-eos}) in the form of both the basic EoS (pressure versus energy density and baryon number density) and an extensive set of crustal composition; appropriate crust EoSs can easily be matched to any core EoS based on their respective values of $J$ and $L$. We shall make available the constant J sequences, the PNM sequences, and also sequences generated by using the $J-L$ correlations from nuclear mass model fits, which pass through the baseline ranges of crust parameters presented in this work.

Finally, we have established how the basic crustal composition varies within experimental and theoretical uncertainties in the CLDM. The model, as mentioned, has some serious drawbacks. The WS approximation breaks down in the highest density regions of the crust - the pasta layers - and shell effects are not taken into account. In upcoming work we shall compare the CLDM results for transition properties and composition with fully microscopic calculations of the crustal matter at the highest densities using a 3D Hartree-Fock method \cite{Newton2009}, in order to establish how much the predictions of the CLDM differ from a more complete model.

\section*{ACKNOWLEDGEMENTS}
We thank the referee for a careful reading of the manuscript and some useful comments which helped improve our analysis. This work is supported in part by the National Aeronautics and Space Administration under grant NNX11AC41G issued through the Science Mission Directorate and the National Science Foundation under grants PHY-0757839 and PHY-1068022 and the Texas Coordinating Board of Higher Education under grant No. 003565-0004-2007.

\section{Appendix: Uniform Nuclear Matter Equations of State}

\subsection{MSL}

The phenomenological modified Skyrme-like (MSL) model \cite{MSL01} will be the form for the uniform, isospin-asymmetric nuclear matter EoS we use for our baseline results. The energy as a function of density $n$ and isospin asymmetry $\delta$ is written down in a form that closely resembles the uniform nuclear matter Skyrme EoS under the Hartree-Fock approach; the advantage of the MSL function is that its free parameters can be easily related to the properties of nuclear matter at saturation, and allow for a smooth variation of those parameters. The MSL EoS is

\be\label{eq:emsl}
   \begin{array}{l l}
	E^{\rm MSL}(n,\delta) &\displaystyle= \frac{\eta}{n}\left(\frac{\hbar^2}{2m_n^*}n_n^{5/3}+\frac{\hbar^2}{2m_p^*}n_p^{5/3}\right)\\
	& \displaystyle + \frac{\alpha}{2}\frac{n}{n_0} + \frac{\beta}{\sigma+1}\frac{n^{\sigma}}{n_0^{\sigma}} + E_{sym}^{loc}(n)\delta^2,
   \end{array}
\ee
\be\label{eq:emsloc}
	E_{sym}^{loc}(n) = (1-y)E_{sym}^{loc}(n_0)\frac{n}{n_0} + yE_{sym}^{loc}(n_0)\left(\frac{n}{n_0}\right)^{\gamma_{\rm sym}},
\ee

\noindent where 
\be\label{effmass1}
{\hbar^2 \over 2m_n^*} = {\hbar^2 \over 2m} + n(C_{\rm eff} + D_{\rm eff} \delta)
\ee
\be\label{effmass2}
{\hbar^2 \over 2m_p^*} = {\hbar^2 \over 2m} + n(C_{\rm eff} - D_{\rm eff} \delta)
\ee

\noindent $\eta = (3/5) (3\pi^2)^{2/3}$, $n_0$ is the saturation density and $\alpha$, $\beta$, $\sigma$, $C_{\rm eff}$, $D_{\rm eff}$, $\gamma_{\rm sym}$, $E_{\rm sym}^{\rm loc}$ and $y$ are free parameters. $C_{\rm eff}$ and $D_{\rm eff}$ are fixed by setting the effective masses at saturation to be $m_{\rm p,0}^* = 0.8m$ and $m_{\rm n,0}^*=0.7m$, where $m$ is the average nucleon mass in free space. We set  $\gamma_{\rm sym}$ = 4/3; $\alpha$, $\beta$, $\sigma$ are set by $K_0$, $n_0$ and $E_0$ while  $E_{\rm sym}^{\rm loc}$ and $y$ control the absolute value of the symmetry energy $J$ and its slope $L$ respectively.

\subsection{BD}

We use as a comparison with MSL a form for the uniform, isospin-asymmetric nuclear matter EoS originally obtained by Bludman and Dover \cite{Bludman1981} and modified by Oyamatsu and Iida \cite{Oyamatsu2003}:

\be\label{eq:ebd}
   \begin{array}{l l}
	E^{\rm BD}(n,\delta) &\displaystyle= \frac{\eta}{n}\left(\frac{\hbar^2}{2m}n_n^{5/3}+\frac{\hbar^2}{2m}n_p^{5/3}\right)\\
	& \displaystyle +(1-\delta^2){v_{\rm s}(n) \over n} + \delta^2 {v_{\rm n}(n) \over n}
   \end{array},
\ee
\be\label{eq:bdvs}
	v_{\rm s} = a_1 n^2 + {a_2 n^3 \over 1 + a_3 n},
\ee
\be\label{eq:bdvn}
	v_{\rm n} = b_1 n^2 + {b_2 n^3 \over 1 + b_3 n}.
\ee

\noindent $v_{\rm s}$ and $v_{\rm n}$ are the potential energy densities for SNM and PNM respectively. The free parameters $a_1, a_2, a_3, b_1, b_2$ are adjusted to obtain the desired values of the saturation properties $K_0$, $n_0$, $E_0$, $J$ and $L$, while $b_3$ which controls the behavior of PNM at high densities is fixed at 1.58632 fm$^3$ as suggested in OI \cite{Oyamatsu2007}; this parameter is analogous to $\gamma_{\rm sym}$ in the MSL model.

\end{document}